\documentclass[11pt]{article}

\usepackage{algorithm} 
\usepackage{algpseudocode} 
\usepackage{amsmath}
\usepackage{amsfonts}
\usepackage{amssymb}
\usepackage[english]{babel}
\usepackage{booktabs}
\usepackage{caption}
\usepackage{cite}
\usepackage{doi}
\usepackage{enumitem}
\usepackage{float}
\usepackage{graphicx}
\usepackage{hyperref}
\usepackage[capitalise]{cleveref}
\usepackage[utf8]{inputenc}
\usepackage{mathptmx}
\usepackage{microtype}
\usepackage{siunitx}
\usepackage{stfloats}
\usepackage{tabularx}
\usepackage{textcomp}
\usepackage{xspace}
\usepackage{color}

\usepackage{subfigure}
\usepackage{tikz}

\usepackage{./template/arxiv}

\usepackage{./input/ao-math-std}
\usepackage{./input/ao-math-graphs}
\usepackage{./input/ao-math-fields}

\newcommand\orcauth[2]{\href{https://orcid.org/#1}
  {\includegraphics[height=0.7em]{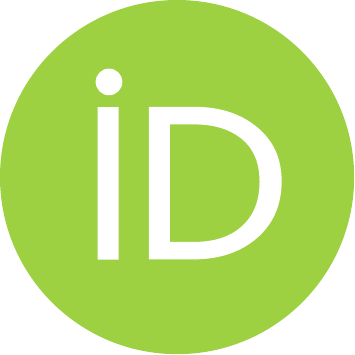}\enspace#2}}

\DeclareSIUnit\mmHg{mmHg}

\title{Connecting continuum poroelasticity with discrete synthetic vascular trees for modeling liver tissue}
\date{}

\author{
    \orcauth{0000-0002-8762-6158}{Adnan Ebrahem}\\
	Institute for Mechanics \\
    Computational Mechanics Group\\
	Technical University of Darmstadt\\
	64287 Darmstadt, Germany\\
    \texttt{adnan.ebrahem@tu-darmstadt.de} \\
\And    
    \orcauth{0000-0003-0276-6029}{Etienne Jessen}\\
	Institute for Mechanics \\
   Computational Mechanics Group\\
	Technical University of Darmstadt\\
	64287 Darmstadt, Germany\\
	\texttt{etienne.jessen@tu-darmstadt.de} \\
\And
    \orcauth{0000-0002-1153-146X}{Marco F.P. ten Eikelder}\\
	Institute for Mechanics \\ Computational Mechanics Group\\
	Technical University of Darmstadt\\
	64287 Darmstadt, Germany\\
	\texttt{marco.eikelder@tu-darmstadt.de} \\
\And
    \orcauth{0000-0002-4524-2180}{Tarun Gangwar}\\
	Discipline of Civil Engineering\\
	Indian Institute of Technology Gandhinagar\\
	Gujarat, India\\
	\texttt{tarun.gangwar@iitgn.ac.in} \\
\And
    \orcauth{0000-0003-3222-9179}{Michał Mika}\\
	Institute for Mechanics \\ Computational Mechanics Group\\
	Technical University of Darmstadt\\
	64287 Darmstadt, Germany\\
	\texttt{michal.mika@tu-darmstadt.de} \\
\And
    \orcauth{0000-0002-9068-6311}{Dominik Schillinger} \\
	Institute for Mechanics \\ Computational Mechanics Group\\
	Technical University of Darmstadt\\
	64287 Darmstadt, Germany\\
	\texttt{dominik.schillinger@tu-darmstadt.de} \\
}

\begin{document}
\maketitle

\begin{abstract}
Computational simulations have the potential to assist in liver resection surgeries by facilitating surgical planning, optimizing resection strategies, and predicting postoperative outcomes.
The modeling of liver tissue across multiple length scales constitutes a significant challenge, primarily due to the multiphysics coupling of mechanical response and perfusion within the complex multiscale vascularization of the organ. In this paper, we present a modeling framework that connects continuum poroelasticity and discrete vascular tree structures to model liver tissue across disparate levels of the perfusion hierarchy. The connection is achieved through a series of modeling decisions, which include source terms in the pressure equation to model inflow from the supplying tree, pressure boundary conditions to model outflow into the draining tree, and contact conditions to model surrounding tissue. We investigate the numerical behaviour of our  framework and apply it to a patient-specific full-scale liver problem that demonstrates its potential to help assess surgical liver resection procedures
\end{abstract}


\newpage
\section{Introduction}
The liver is a highly vascularized organ serving several physiological functions, such as metabolism of nutrients and drugs, detoxification, bile production, and hormone regulation \cite{RefDebbautDiss}. A liver resection, or hepatectomy, is a common surgical procedure to remove part of the liver, mostly due to a (pre-)cancerous or benign tumor.  The liver's complex vasculature makes it challenging to predict the impact of a surgical resection accurately. Computational models can help predict the impact on blood perfusion and determine the amount of liver tissue that can be removed safely while functionality is maintained.

Patient-specific modeling of surgical liver resection requires the adequate modeling of the liver's hierarchical vasculature. 
Blood is supplied to the liver through the hepatic artery, that comes from the heart 
and the portal vein. 
These two vessels branch into vessels of smaller diameter, forming vascular trees, which supply the liver parenchyma with blood. After passing the liver microcirculation, 
blood is recollected via smaller and then larger vessels of the hepatic vein and goes back to the heart \cite{RefDebbautDiss}. Identifying multiscale vascular trees in vivo through imaging is impossible due to limited resolution. They must therefore be generated synthetically with the help of a computer. The best-known generation method is constrained constructive optimization (CCO) \cite{Schreiner1,Schreiner2}. Its core is a local optimization approach, directly based on Murray’s minimization principles \cite{RefMurray}.
We recently extended the CCO approach such that a tree can be found that is optimal both in (global) geometry
and topology \cite{RefEtienne}. Optimizing the geometry is cast into a nonlinear optimization problem, which
allows the investigation of various possible goal functions and constraints \cite{RefEtienneMurray}. The resulting synthetic trees showed good agreement with real trees of a human liver characterized experimentally from corrosion casts. We recently extended our technology to the simultaneous generation of multiple supplying and draining trees \cite{Jessen3}.


Blood perfusion is closely linked to tissue deformation, and including tissue deformation enhances the predictive capability of the liver tissue model. 
Unfortunately, synthetic vascular trees, which are largely based on optimization principles, do not offer a direct link to be coupled with tissue mechanics and deformation. 
One solution is to resort to homogenization and the theory of poromechanics, replacing the complex heterogeneous medium by a fictitious homogeneous medium with equivalent macroscale behaviour.

Many existing studies based on poromechanics considered either perfusion models \cite{RefLiverPerfusion,RefLiverPerfusion1,RefLiverPerfusion2,RefLiverPerfusion3,RefLiverPerfusion4,RefRicken2} or tissue deformation \cite{RefLivershear}.
In \cite{RefLiverPerfusion}, a perfusion system is proposed that is decomposed into compartment models, each valid at a different scale, to describe blood flow in the human liver. 
In \cite{RefStoter}, an approach to model perfusion in a patient-specific human liver is based on a diffuse interface method that couples porous-medium-type flows.
To date, there are only a few studies that consider liver perfusion coupled to tissue deformation. 
In \cite{RefLivershear}, a porohyperviscoelastic model is used to predict shear waves in pressurized soft liver tissues.
In \cite{RefRicken}, a multiphasic model was developed to describe transport phenomena and perfusion metabolism in the liver, where idealized two-dimensional liver structures, representing liver lobules, are considered. 

In this paper, we show that the two modeling approaches, i.e., discrete synthetic vascular trees and continuum poroelasticity, can be synergistically combined. We demonstrate that the resulting framework has the potential to support the assessment of surgical resection procedures by simulating the impact on the liver's perfusion characteristics.
Our paper is organized as follows. In Section \ref{sec:2} we provide the poroelastic model in terms of a two-phase pressure-displacement formulation with incompressible constituents. Section \ref{sec:3} reviews our method for synthetically generating vascular trees based on mathematical optimization. In Section \ref{sec:4}, we connect our continuum poroelastic and discrete vascular tree models via suitable interface assumptions on geometry and boundary conditions. Additionally, we derive the weak formulation of the poroelastic model. In Section \ref{sec:5}, we first discuss the characteristic behaviour of the poroelastic model via a two-dimensional test problem, and then apply it to simulate a three-dimensional model of a liver resection. Section \ref{sec:6} closes with a discussion and an outlook.

\section{Continuum poroelastic model}\label{sec:2}
In this section, we briefly review poroelasticity at large strains and provide the balance laws and the constitutive laws of the poroelastic model that we will use in the following.

\subsection{Preliminaries and kinematics}\label{subsec: prelim and kinem}
Classical poromechanics is rooted in continuum mixture theory \cite{RefTruesdell}. Continuum mixture theory is a general mathematical theory that provides a framework for deriving (simplified) continuum mechanics models for a large number of multi-physics problems. For an extensive review on poromechanics we refer to \cite{RefBookCoussy,RefBookEhlers,RefBookBoer}, and note a number of important theoretical and numerical studies in the field of poromechanics \cite{RefLungTree,RefChapelle,RefMarkert,RefPhysAppl,Refheart1,RefWallPorous,RefDormieux2}.

The core principle in poromechanics is that the porous material is composed of multiple constituent bodies that simultaneously occupy a common region in space. In this work, we rely on the common assumption that the Lagrangian configuration of the constituent bodies coincide. This means that we work with a single Lagrangian description. Hence, the spatial position (motion) of a particle is given by the (invertible) deformation map:
\begin{equation}
    \boldsymbol{x} = \boldsymbol{\chi}(\boldsymbol{X},t),
\end{equation}
where $\boldsymbol{X} \in \Omega_0$ denotes the Lagrangian position, $\boldsymbol{x}\in \Omega$ the spatial position, and $t$ the time. Here $\Omega_0$ and $\Omega$ are the reference and current domain of the mixture, respectively. We use the standard notation for the displacement of the mixture, i.e. $\boldsymbol{u}=\boldsymbol{x}-\boldsymbol{X}$. Furthermore, we denote the Lagrangian velocity as $\boldsymbol{v}=\dot{\boldsymbol{u}}$, where the dot represents the material derivative. We introduce the following kinematic quantities: 
\begin{subequations}
\begin{align}
         \boldsymbol{F} =&~ \boldsymbol{1} + \nabla \boldsymbol{u}, \\
         J =&~ {\rm det} \; \boldsymbol{F}, \\
         \boldsymbol{C} =&~ \boldsymbol{F}^T \boldsymbol{F},\\
         \boldsymbol{E} =&~ \frac{1}{2}(\boldsymbol{F}^T \boldsymbol{F}-\boldsymbol{1}),
\end{align}
\end{subequations}
where $\boldsymbol{F}$ is the deformation gradient, $J$ its determinant, $\boldsymbol{C}$ the right Cauchy-Green tensor, and $\boldsymbol{E}$ the Green-Lagrange strain tensor.

\begin{figure}[ht]
    \begin{tikzpicture}
   \node[] (pic) at (0,0) {\includegraphics[width=135mm]{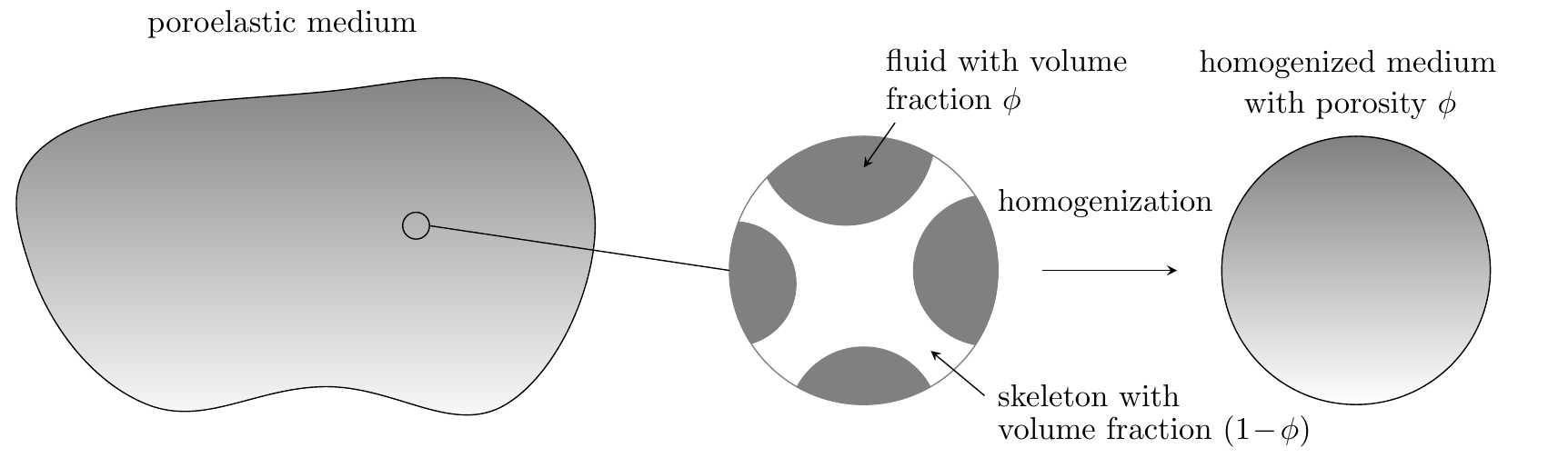}};
     \end{tikzpicture}
    \caption{Continuum (homogenized) poroelastic mixture consisting of a skeleton and fluid constituent.}
    \label{fig:homogenization}
\end{figure}

In this work, we consider a heterogeneous mixture composed of a single fluid and a single solid constituent, where superscripts $f$ and $s$ refer to quantities associated with the fluid and the skeleton phase, respectively.
We denote the volume fraction of the fluid and solid (skeleton) constituent respectively as $\phi^f$ and $\phi^s$. 
Since the skeleton is a deformable macroscopic structure, its deformation changes the structure of its pores. 
As a consequence, the volume fractions are time-dependent (and obviously space-dependent), i.e. $\phi^f=\phi^f(\boldsymbol{x},t)$ and $\phi^s=\phi^s(\boldsymbol{x},t)$.
We assume that void spaces are absent, i.e. 
  \begin{align}
  \phi^f(\boldsymbol{x},t) + \phi^s(\boldsymbol{x},t) = 1,
  \end{align}
for all $\boldsymbol{x}\in \Omega$ and $t\geq 0$.
As a consequence, the composition can be described by the porosity $\phi=\phi(\boldsymbol{x},t)$:
\begin{subequations}
  \begin{align}
    \phi^f &= \phi, \\
    \phi^s &= 1 - \phi.
  \end{align}
\end{subequations}
At the macroscopic level, the solid-fluid mixture is typically considered a homogenized medium. We visualize our model in \cref{fig:homogenization}.

The partial mass densities of the fluid and solid constituents denote $\tilde{\rho}^f=\tilde{\rho}^f(\boldsymbol{x},t)$ and $\tilde{\rho}^s=\tilde{\rho}^s(\boldsymbol{x},t)$, respectively. These densities represent the mass of the associated constituent per infinitesimal mixture volume. The partial mass densities may be decomposed as:
\begin{subequations}
    \begin{align}
      \tilde{\rho}^f=&~ \rho^f \phi,\\
      \tilde{\rho}^s=&~ \rho^s (1-\phi).
    \end{align}
\end{subequations}
In this paper, we assume that both constituents are incompressible, i.e.
\begin{subequations}
    \begin{align}
    \rho^f =&~ \text{const},\\
    \rho^s =&~ \text{const}.
    \end{align}
\end{subequations}
The assumption of incompressible constituents is common in biomechanics, since the fluid pressure and solid stresses are typically negligible in comparison to the bulk modulus of the material \cite{RefPhysAppl}.

\subsection{Balance laws and constitutive equations}

In agreement with the continuum theory of mixtures, each constituent may be considered in isolation and its motion involves terms that model the interaction with the other constituents. The motion of the mixture is then a consequence of the individual evolution equations. In the scope of this work, we focus directly on the evolution equations relevant for the final poroelastic model, 
assuming quasi-static conditions.

The balance of fluid mass takes the form:
\begin{align}
 \partial_t (\rho^f \phi) + \nabla \cdot \left( \rho^f \phi \boldsymbol{v}^f \right) =\rho^f \theta,
\end{align}
where $\boldsymbol{v}^f$ denote the velocity of the fluid and $\theta$ describe a mass source term $(\theta \geq 0)$ or a sink term $(\theta \leq 0)$ \cite{Refheart1}.
Next, we introduce the added mass quantity $m$ and the perfusion velocity $\boldsymbol{w}$ as:
\begin{subequations}
    \begin{align}
    m =&~ \tilde{\rho}^f J - \tilde{\rho}^f_0,\\
    \tilde{\rho}^f_0 =&~ \rho^f \phi_0,\\
    \boldsymbol{w} = &~ \phi (\boldsymbol{v}^f-\boldsymbol{v}),
\end{align}
\end{subequations}
where $\phi_0=\phi_0(\boldsymbol{X}) = \phi(\boldsymbol{\chi}^{-1}(\boldsymbol{X},0),0)$  represents the porosity in the reference configuration. The added mass $m$ represents the variation in fluid mass content per unit volume of the undeformed skeleton. A straightforward calculation reveals that the evolution of the added mass is given by:
\begin{align}
    \frac{1}{J} \dot{m} + \nabla \cdot \left(\rho^f \boldsymbol{w} \right) = \rho^f \theta.
    \label{eqn:massbalance3}
\end{align}
We relate the perfusion velocity $\boldsymbol{w}$ to the fluid pressure $p$ in the pores using Darcy's law:
\begin{align}\label{eqn:darcy}
     \boldsymbol{w} = - \frac{\boldsymbol{k}}{\eta} \nabla p,
\end{align}
where the quantity $\boldsymbol{k}$ describes the symmetric second order permeability tensor of the mixture and $\eta$ is the dynamic viscosity. We restrict ourselves to the isotropic case, i.e $\boldsymbol{k}=k \boldsymbol{1}$ with $k = {\rm const}$ and rewrite $K = \frac{k}{\eta}$. We note that Darcy's law as constitutive relation is a standard choice in the literature. Considering the steady-state case and substitution of (\ref{eqn:darcy}) into (\ref{eqn:massbalance3}) provides
\begin{subequations}
   \label{eqn:finalflow}
    \begin{align}
      - K \nabla^2 p &= \theta \; \;\; \, \text{in} \; \Omega, \label{eqn:finalflow_a}\\
      - K \left( \boldsymbol{F}^{-T} \nabla_0\right) \boldsymbol{F}^{-T}\nabla_0 p &= \theta \; \;\; \, \text{in} \; \Omega_0,
    \label{eqn:finalflow_b}
    \end{align}
\end{subequations}
where we have used the pull-back operation $\nabla = \boldsymbol{F}^{-T}\nabla_0$ for the mapping of (\ref{eqn:finalflow_a}) to the reference configuration $\Omega_0$ with $\nabla_0$ denoting the material gradient.
The balance of momentum in the actual and reference configuration may then be written as:
\begin{subequations}
\begin{align}
    \nabla \cdot \boldsymbol{\sigma} =&~ \boldsymbol{0}  \; \;\; \, \text{in} \; \Omega, \\
    \nabla_0 \cdot \left(\boldsymbol{F} \boldsymbol{S} \right) =&~ \boldsymbol{0}  \; \;\; \, \text{in} \; \Omega_0,
\end{align}    
\end{subequations}
where we have assumed the absence of body forces. Here $\boldsymbol{\sigma} = J^{-1}\boldsymbol{F} \boldsymbol{S} \boldsymbol{F}^T$ is the Cauchy stress tensor for the complete medium and $\boldsymbol{S}$ represents the second Piola-Kirchhoff stress tensor. To consider the role of the interstitial fluid, we introduce the effective stress, also referred to as Terzaghi decomposition:
\begin{align}
    \boldsymbol{\sigma} = \boldsymbol{\sigma}' - p\boldsymbol{1},
\end{align}
where $\boldsymbol{\sigma}'$ denotes the effective stress. The deformation of the skeleton is now determined by the effective stress $\boldsymbol{\sigma}'$. We choose the following constitutive relations:
\begin{subequations}
\begin{align}
        \boldsymbol{S}=&~\frac{\partial \Psi^s(\boldsymbol{E},J^s)}{\partial \boldsymbol{E}}-pJ\boldsymbol{C}^{-1},\\ p=&~-\frac{\partial \Psi^s(\boldsymbol{E},J^s)}{\partial J^s},
\end{align}
\end{subequations}
where $\Psi^s=\Psi^s(\boldsymbol{E},J^s)$ is the Helmholtz free energy density and $J^s = J(1-\phi)$ the Jacobian weighted by the volume fraction of the skeleton phase \cite{RefWallPorous}. The introduced constitutive equations arise from thermodynamic principles on a macroscopic scale. For a review on poroelasticity from the microscopic perspective and the derivation of constitutive relations by means of a micro-macro approach, we refer to \cite{RefDormieux2}.

To close the system of equations, the Helmholtz free energy needs to be selected. We choose to work with a free energy that decomposes as
\begin{align}
\Psi^s(\boldsymbol{E},J^s) = \Psi^{\text{skel}}(\boldsymbol{E}) +   \Psi^{\text{vol}}(J^s), 
\end{align}
where $\Psi^{\text{skel}}(\boldsymbol{E})$ is the hyperelastic potential of the skeleton and $\Psi^{\text{vol}}(\boldsymbol{E})$ accounts for macroscopic volume change due to interstitial fluid pressure. 
In this work, we employ a hyperelastic material model of Neo-Hookean type for the skeleton which can be expressed in terms of the first and third invariant of the right Cauchy-Green tensor: 
\begin{equation}
\Psi^\text{skel} = \frac{1}{8}\lambda \text{ln}^2(I_3)+\frac{1}{2}\mu[I_1 - 3 - \text{ln}(I_3)],
\end{equation}
with the invariants $I_1 = \text{tr}\; \boldsymbol{C}$ and $I_3 = \text{det}\; \boldsymbol{C}$. The coefficients $\lambda$ and $\mu$ describe the Lamé parameters. 
For the volumetric contribution of the free energy function, we choose 
\begin{align}
    \Psi^\text{vol} = \kappa \left( \frac{J^s}{1-\phi_0} - 1 - \text{ln}\left( \frac{J^s}{1-\phi_0}\right) \right),
    \label{eqn:volumetric}
\end{align}
with $\kappa = E/(3(1-2\nu))$ denoting the bulk modulus of the skeleton \cite{RefWallPorous}.
With that choice, the constitutive equations can be rewritten as
\begin{align}
   \boldsymbol{S}&=2\frac{\partial \Psi^\text{skel}}{\partial \boldsymbol{C}}-pJ\boldsymbol{C}^{-1},\\
    p&=-\frac{\partial \Psi^\text{vol}}{\partial J^s}. \label{eqn:pressureconsitituive}
\end{align}
Equation \eqref{eqn:pressureconsitituive} relates $J^s$ to $p$, and thus the porosity $\phi$ to the fluid pressure $p$. Inserting (\ref{eqn:volumetric}) into (\ref{eqn:pressureconsitituive}) we obtain
\begin{align}
    p = \kappa\left( - \frac{1}{1-\phi_0} + \frac{1}{J^s}\right).
    \label{eqn:pressureporosity}
\end{align}
In summary, the poroelastic model is given by the following system of equations in the reference configuration $\Omega_0$:
\begin{subequations}
 \label{eqn:coupledsystem}
 \begin{align}
    \nabla_0 \cdot (\boldsymbol{F} \boldsymbol{S}) &= \boldsymbol{0} \;\; \;\;\;\;\;\;\;  \, \text{in} \; \Omega_0, \label{eqn:momentumequation} \\
    - K \left( \boldsymbol{F}^{-T} \nabla_0\right) \boldsymbol{F}^{-T}\nabla_0 p &= \theta \;\; \;\; \;\; \;\;\; \text{in} \; \Omega_0, \label{eqn:masscoupledsystem}\\
       \boldsymbol{S}&=2\frac{\partial \Psi^\text{skel}}{\partial \boldsymbol{C}}-pJ\boldsymbol{C}^{-1},\\
    p&=-\frac{\partial \Psi^\text{vol}}{\partial J^s},
	\end{align}
\end{subequations}
where the displacement $\boldsymbol{u}$ and the fluid pressure $p$ are the two primary variables. The system needs to be complemented with suitable boundary conditions. We will specify these in Section \ref{sec:4}.

\section{Discrete vascular tree model}\label{sec:3}
We now briefly describe the model assumptions and generation of vascular trees based on a set of physiological constraints, where we closely follow our work on synthetic vascular trees \cite{RefEtienne,Jessen3}.

\subsection{Mathematical formulation}
We describe each vascular tree as a directed graph $\mathbb{T} = \left(\mathbb{V}, \mathbb{A}\right)$ with nodes $u \in \mathbb{V}$ and segments $a \in \mathbb{A}$.
Each segment $a = uv$ approximates a \emph{vessel} as a rigid and straight cylindrical tube defined by the geometric locations of nodes $x_u$ and $x_v$, length $\ell_a = ||x_u - x_v||$, volumetric flow $Q_a$ and radius $r_a$.
The proximal node of the single root segment is the \emph{root} $x_0$, and the distal nodes of each terminal segment are the \emph{leaves} $v \in \mathbb{L}$.
We approximate blood as an incompressible, homogeneous Newtonian fluid and assume laminar flow through each vessel of the tree.
The hydrodynamic resistance $R_a$ of each segment $a$ can be described by Poiseuille's law:
\begin{equation}
  R_a = \frac{8 \eta}{\pi} \frac{\ell_a}{r_a^4} \quad\forall a \in \mathbb{A},
\end{equation}
where $\eta$ is the dynamic viscosity of blood, set to $\SI{3.6}{cP}$.
The pressure drop across a segment follows then with
\begin{equation}
  \Delta p_a = R_a Q_a \quad\forall a \in \mathbb{A}.
\end{equation}
At branching nodes, the relationship between parent and child segments obeys Murray's law \cite{RefMurray}, defined by 
\begin{equation}
  r_{uv}^3 = \sum_{vw \in \mathbb{A}} r_{vw}^3 \quad\forall v \in \mathbb{V} \setminus \mathbb{L}.
\end{equation}
Each tree is perfused at steady-state by a given perfusion flow $Q_\text{perf}$. 
We assume a homogeneous flow distribution to all $N$ leaves with a terminal flow $Q_\text{term} = Q_\text{perf}/N$ and use Kirchhoff's law to compute the flow values of the branching nodes with 
\begin{equation}
Q_{uv} = \sum_{vw \in \mathbb{A}}  Q_{vw} \quad\forall v \in \mathbb{V} \setminus \left({0} \cup \mathbb{L}\right).
\end{equation}
The trees are generated to obey scaling relations based on minimizing the total power, which consists of the power to maintain blood inside the vessels $P_\text{vol}$ and the (viscous) power to move blood through vessels $P_\text{vis}$. 
The total cost of a vascular tree thus is defined with
\begin{equation}
    f_\mathbb{T} = P_\text{vol} + P_\text{vis} = \sum_{a \in \mathbb{A}} m_b\pi\ell_a r_a^2 +\frac{8 \eta}{\pi} \frac{\ell_a}{r_a^4}Q_a^2,
\end{equation}
where $m_b$ is the metabolic demand factor of blood, which we set to \SI{0.6}{\uW\per\cubic\mm}.

\subsection{Algorithmic solution approach}
Our aim is to generate a set of one supplying and one draining tree inside the liver, which obey these goals and constraints and are optimal both in topology and geometry. 
Using the framework described in \cite{RefEtienne} for each tree, we start by generating $N$ terminal nodes $\bar{x}$ inside the perfusion volume and connect them to the manually set root position.
From this initial (fan) shape, new topologies are explored by swapping segments. 
A \emph{swap} detaches a node from its parent and connects it with another existing segment.
Afterwards, the global geometry (the positions of all branching nodes) is optimized by solving a nonlinear optimization problem (NLP).
The newly created topology is accepted based on a Simulated Annealing approach, and new swaps are created until the topologies of both trees converge against a local minimum.
If the resulting swap creates an intersection between the supplying and draining tree, we always reject it.

For the global geometry optimization, we include the nodal positions $x$, the length $\ell$ and the radii $r$ of all segments inside the vector of optimization variables $y = (x, \ell, r)$.
We introduce physical lower bounds $\ell^-, r^-$ and numerical upper bounds $\ell^+, r^+$.
The best geometry is then found in 
\begin{equation}
    Y = \mathbb{R}^{3|\mathbb{V}|} \times [\ell^-, \ell^+]^\mathbb{A} \times [r^-, r^+]^\mathbb{A}
\end{equation}
and our NLP reads:
\begin{align}\label{nlp-murray}
  \min_{y \in Y} \quad
  & \sum_{a \in \mathbb{A}} m_b\pi\ell_a r_a^2 + 8\eta/\pi Q_a^2\ell_a/r_a^4\\
  \text{s.t.}\quad
  \label{eq:nlp-fix-x}
  &0 = x_u - \bar{x}_u, & u &\in \mathbb{V}_0 \cup \mathbb{L}\\
  \label{eq:nlp-length}
  &0 = \ell_{uv}^2 - ||x_u - x_v||^2, & uv &\in \mathbb{A}\\
  \label{eq:nlp-murray-law}
  &0 = r_{uv}^3 - \sum_{vw \in \mathbb{A}} r_{vw}^3 & v &\in \mathbb{V} \setminus (0 \cup \mathbb{L})
\end{align}
(\ref{eq:nlp-fix-x}) fixes the position of terminal nodes, (\ref{eq:nlp-length}) ensures consistency between nodal positions and segment length and (\ref{eq:nlp-murray-law}) enforces Murray's law. 
After the trees are successfully generated, all nodal positions $x$ are fixed. At each segment $a$, we can now directly retrieve the length $\ell_a$, the radius $r_a$, and the volumetric flow $Q_a$. Furthermore, the mean velocity $\bar{v}_a$ through each segment $a$ can be easily computed with
\begin{equation}
    \bar{v}_a = \frac{Q_a}{\pi r_a^2}.
\end{equation}

\section{A phenomenological modeling framework for tissue perfusion}\label{sec:4}
In this section, we describe the coupling of the vessel trees to the poroelastic model derived in the previous section. First, we describe the interaction of the poroelastic domain with surrounding tissues by nonlinear displacement boundary conditions. We then introduce modeling assumptions in terms of source terms for the inlets and boundary conditions for the outlets to enable the perfusion of the poroelastic domain. We close this section by deriving the weak form of the coupled problem for the purpose of finite element discretizations.

\subsection{Modeling the interaction with surrounding tissues}\label{subsec: Modeling the interaction with surrounding tissues}
To arrive at a closed boundary value problem, we need to complement the system \eqref{eqn:coupledsystem} by appropriate boundary conditions. In our application case, we would like to take into account the interaction of the liver with surrounding organs, with which the liver is continuously in contact. 
\begin{figure}[ht]
\vspace{-1.0cm}
\centering
 \begin{tikzpicture}
   \node[] (pic) at (0,0) {\includegraphics[width=85mm]{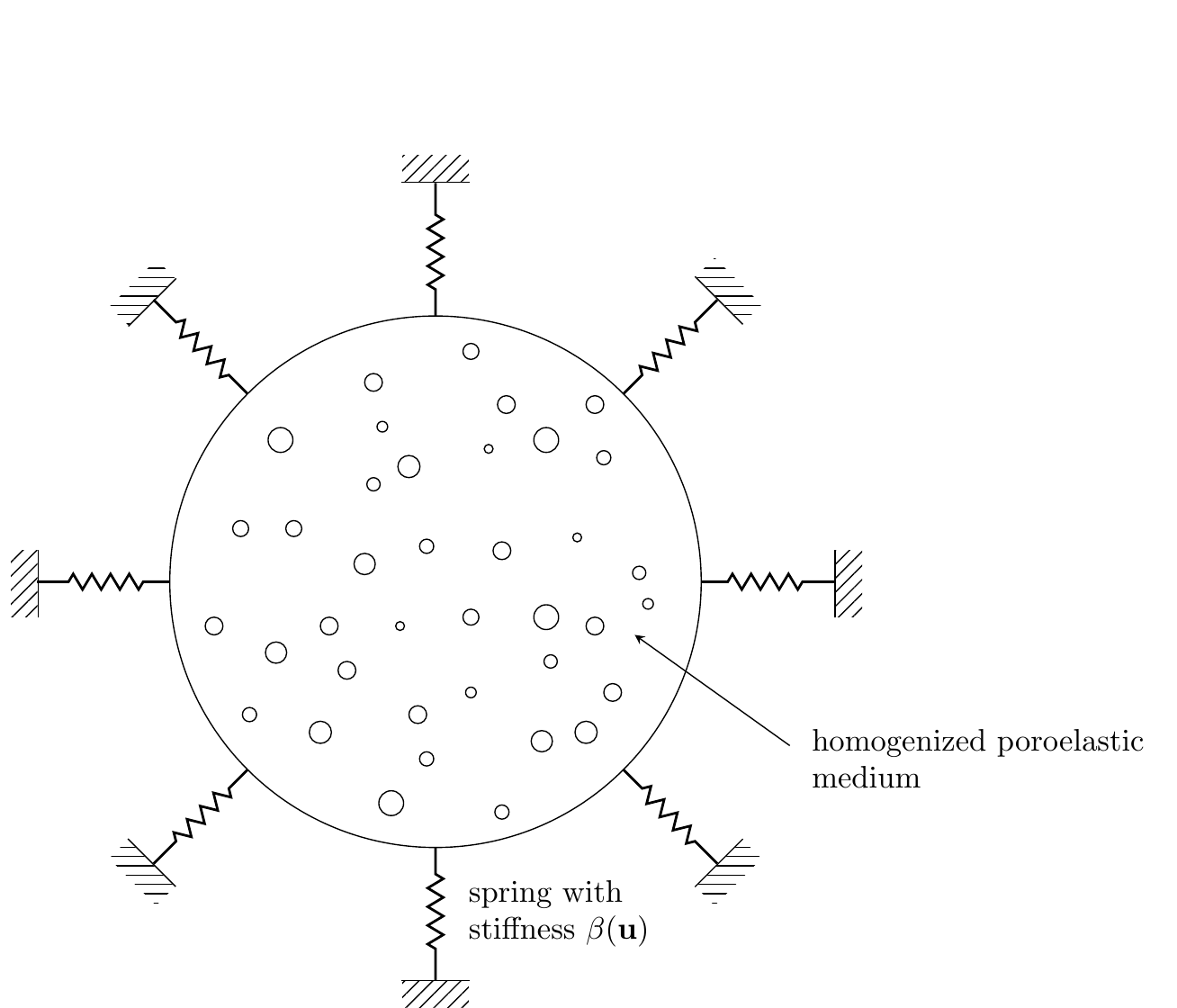}};
    \node[] (pic) at (5.5,1) {\includegraphics[width=60mm]{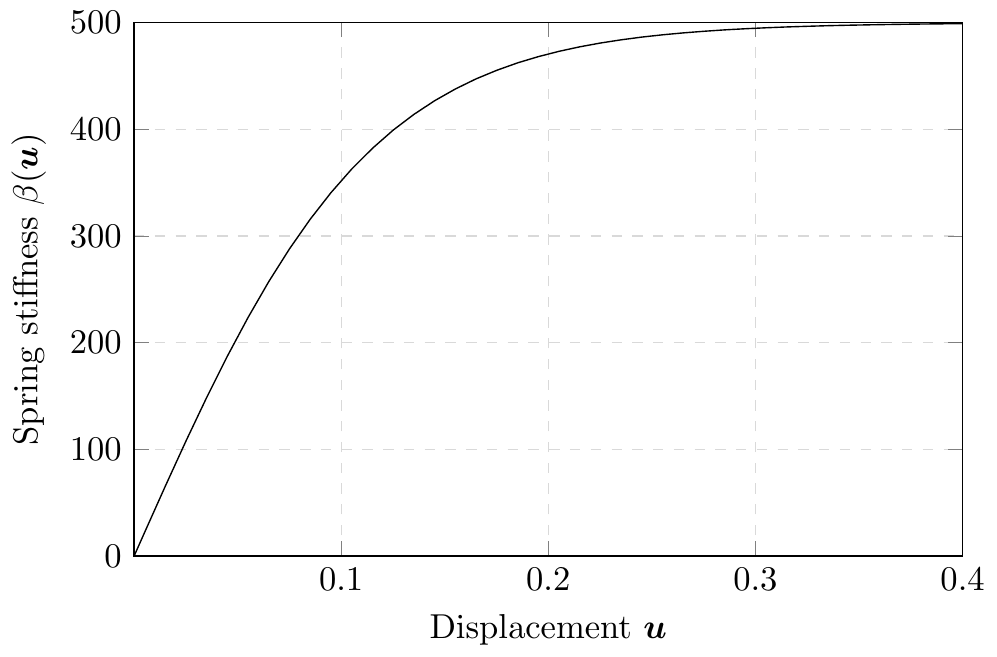}};
     \end{tikzpicture}
\caption{Modeling the resistance of surrounding tissues. The spring stiffness $\beta$ is a function of the displacements $\boldsymbol{u}$.}
\label{fig:springcontact}
\end{figure}
Motivated by a penalty approach known from contact mechanics, we model the resistance of the surrounding organs by adding the following contribution:
\begin{align}
    W_\text{c}(\boldsymbol{u}) =  \beta \boldsymbol{u} 
\end{align}
supported on $\Gamma_{\rm outer}$ to the left-hand side the balance of momentum. This term can be interpreted to mimic the effect of nonlinear springs at the outer boundary as illustrated in Fig. \ref{fig:springcontact}, where $\beta$ corresponds to the spring stiffness. In hyperelastic tissue-like materials, the stiffness changes with the deformation. We therefore model $\beta$ as a function of the displacements $\boldsymbol{u}$:
\begin{align}
    \beta(\boldsymbol{u}) = \frac{2\alpha}{1+e^{-c\boldsymbol{u}}}-\alpha,
    \label{eqn:stiffnesscontact}
\end{align}
in which $\alpha$ corresponds to the maximum value of the spring stiffness and $c$ is the steepness of the curve. We choose $c = 15$ for all computations. Analogous to nonlinear springs, the stiffness saturates towards a constant value with increasing displacement. 

\subsection{Augmenting the poroelastic model with discrete tree feature}
\begin{figure}[t]
\centering
   \begin{tikzpicture}
   \node[] (pic) at (0,0) {\includegraphics[width=135mm]{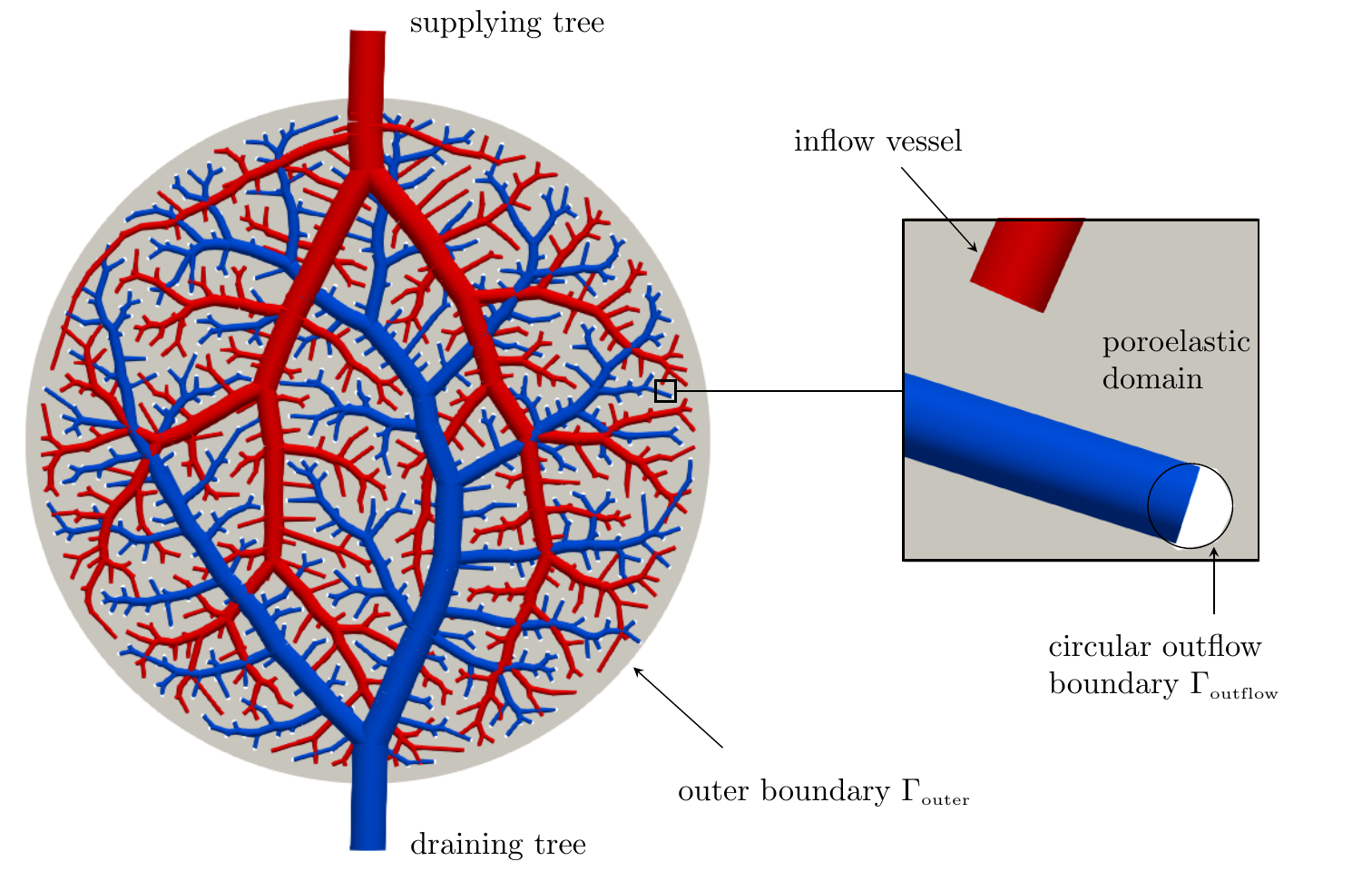}};
     \end{tikzpicture}
 	\caption{Coupling of the vascular trees with the porous medium. The red tree represents the vessels of the supplying tree while the blue ones are the vessels of the draining tree. Circular voids represent the interface area between the continuum model and the discrete draining tree.}
  \label{fig:coupledvesseltree}
\end{figure}

The poroelastic domain, representing the tissue, is supplied with fluid from the vessels of an supplying tree and returns fluid through the vessels of a draining tree (see Fig. \ref{fig:coupledvesseltree}). Therefore, the poroelastic domain can be interpreted as a connector between the supplying and draining trees. 
We now address the question how to connect the poroelastic model to the vessel trees by specifying appropriate boundary conditions to induce flow from the inlets to the outlets.

To bridge the gap between the macroscopic (homogenized) medium and the discretely resolved levels of the vascular tree, we assume circular areas (or spherical areas in 3D), whose radii are of the same order as the radii of the vessels at the terminal vessel points (see Fig. \ref{fig:coupledvesseltree}). We cannot model the physiological mechanisms in these areas directly, and therefore depict them as void. In the following, we describe corresponding modeling assumptions in terms of source terms for the inlets and boundary conditions for the outlets.  

\subsubsection{Bell-shaped source terms to model flow from the supplying tree}
We induce flow from the discrete supplying tree into the poroelastic domain through the source quantity $\theta$ in the mass conservation equation \eqref{eqn:masscoupledsystem} by a summation over all $n$ terminal vessels of the supplying tree:
\begin{align}
    \label{eqn:summationgauss}
    \theta = \sum_{i=1}^{n} \theta_i,
\end{align}
where $\theta_i$ refers to the source term of the $i$-th terminal vessel of the supplying tree. We transfer the volumetric flow from each terminal vessel $i$ of the supplying tree into the source term in the mass conservation equation \eqref{eqn:masscoupledsystem} in form of a bell-shaped distribution:
   \begin{align}
   \label{eqn:gauss}
    \theta_i(\boldsymbol{x}) &= \gamma_i \; \text{exp}\left( \frac{-\lVert \left( \boldsymbol{x} - \boldsymbol{x}_i\right) \rVert_2^2}{(br_i)^2} \right),
\end{align}
where $\gamma_i$ is the amplitude of the $i$-th function, $\lVert \cdot \rVert_2$ is the Euclidean norm, $\boldsymbol{x}_i$ is the position vector of the $i$-th inlet terminal point, $r_i$ is the radius of the corresponding $i$-th inlet terminal vessel and $b$ is a scaling factor of the radius. The radii $r_i$, the locations $\boldsymbol{x}_i$, and the volumetric flow $Q_i = \int \theta_i {\rm d}\mathbf{x}$ are extracted from the supplying vascular tree data described in section \ref{sec:3}. To satisfy conservation of mass, we require that the total flow that enters the domain, $\sum_i Q_i$, matches to the total flow that leaves the draining tree.

The bell-shaped function possesses several advantages that justifies this choice. 
The symmetric, smooth and continuous nature of the bell-shaped function distributes the inflow, modeling the effect of the interface area that is not represented in the discrete and continuum models. It allows for a simple and effective control of  the overall shape and magnitude of the inflow profile. 
The bell-shaped function also has a well-defined peak that represents the highest flow rate. Moreover, the bell-shaped function can be employed to simulate the spread of the fluid as it enters the domain. By adjusting the amplitude of the bell-shaped function, the magnitude of the inflow can be controlled. The bell-shaped function has a simple mathematical form and an analytical solution that allows for efficient and accurate computation in numerical simulations. In particular, the amplitude $\gamma_i$ is determined from the known $i$-th volumetric flow $Q_i$ via $\gamma_i = Q_i/(\pi b^2 r_i^2)$ (two dimensions) and $\gamma_i = Q_i/((\pi b^2 r_i^2)^{\frac{3}{2}})$ (three dimensions).  

\subsubsection{Boundary conditions to model flow into the draining tree}

We induce flow from the poroelastic domain into the discrete draining tree by imposing Dirichlet boundary conditions for the pressure at the circular boundaries of these void areas (denoted by $\Gamma_\text{outflow}$). The radii and locations of the terminal outlet points are extracted from the draining vascular tree data described in section \ref{sec:3}. It is convenient to set a reference pressure level of $p = 0$ here. 
To guarantee the conservation of mass, we model the outer boundary (denoted by $\Gamma_\text{outer}$) of the domain as impermeable by inducing the Neumann boundary condition
\begin{equation}
    \nabla p \cdot \mathbf{n} = 0,
\end{equation}
which guarantees that no fluid is leaving  the poroelastic domain through its outer boundary. 

\subsection{Weak formulation and discretization}
\label{sec:fem}
We utilize the standard finite element method \cite{RefHughes} for the discretization of the poroelastic model in the Lagrangian description \eqref{eqn:coupledsystem} augmented with the interaction term presented in \cref{subsec: Modeling the interaction with surrounding tissues}. Multiplication of the momentum equation (\ref{eqn:momentumequation}) with discrete test function $\boldsymbol{v}_h$ and the pressure equation (\ref{eqn:masscoupledsystem}) with discrete test function $q_h$, and subsequently integrating over the reference domain $\Omega_0$, and applying integration by parts leads to the weak statement: Find $\boldsymbol{u} \in \mathcal{V}_{h}$ and $p \in \mathcal{W}_{h,0}$ such that for all $\boldsymbol{v}_h \in \mathcal{V}_{h}$ and $q_h \in \mathcal{W}_{h,0}$:
\begin{align}
    \int\displaylimits_{\Omega_0}(\boldsymbol{F} \boldsymbol{S}) : \nabla_0\boldsymbol{v}_h \, \text{d}\Omega_0 + \int\displaylimits_{\Gamma_{{\rm outer}}} \beta \boldsymbol{u} \cdot \boldsymbol{v}_h  \, \text{d}\Gamma_{{\rm outer}} &= \textbf{0}, \label{eqn:weakproblemsolid_reference} \\
    \int\displaylimits_{\Omega_0} \left( K \mathbf{F}^{-T} \nabla_0 p \right) \cdot \left(\mathbf{F}^{-T} \nabla_0 q_h\right) \,  \text{d}\Omega_0 + \int\displaylimits_{\Gamma_{{\rm outer}}} \left(K\nabla_0 p \cdot \mathbf{N}\right) q_h \, \text{d}\Gamma_{{\rm outer}}  &= \int\displaylimits_{\Omega_0} \theta q_h \, \text{d}\Omega_0,\label{eqn:fluidproblem1_reference}
\end{align}
where $\mathbf{N}$ is the normal vector in the reference configuration. The discrete function spaces $\mathcal{V}_{h}$ and $\mathcal{W}_{h,0}$ consist of linear and quadratic Lagrange basis functions of degree $P=1$ and $P=2$, and are applied to discretize the displacements and the pressure, respectively \cite{RefReducedDarcy}. Homogeneous Dirichlet boundary conditions on $\Gamma_\text{outflow}$ are strongly enforced in $\mathcal{W}_{h,0}$:
\begin{align}
    \mathcal{W}_{h,0} &=  \{ q_h \in \mathcal{W}_{h}: q_h = 0 \, \, \text{on} \, \Gamma_\text{outflow}  \}, 
\end{align}
with $\mathcal{W}_{h}$ being the unrestricted function space for the pressure. We implemented the framework in FEniCS, where we utilized a standard Newton-Raphson method, the iterative solver GMRES and the preconditioner Hypre\_Euclid \cite{RefFenics}. 

Equation (\ref{eqn:masscoupledsystem}), also referred to as the reduced Darcy formulation \cite{RefReducedDarcy}, is solely written in terms of pressure, as we have eliminated the velocity upon substituting (\ref{eqn:darcy}) into (\ref{eqn:massbalance3}). Not substituting (\ref{eqn:darcy}) into (\ref{eqn:massbalance3}) leads to a two-field formulation (velocity and pressure), also referred to as the full Darcy system in literature \cite{RefReducedDarcy}. In that case, the poroelastic equations have a saddle point structure and the discrete pressure and velocity spaces must therefore satisfy the inf-sup condition \cite{RefBabuska,RefBrezzi}. One stable combination of mixed finite element pairs is for example a Taylor-Hood element with a pressure approximation that is one order lower than the one for the velocity. 
Disadvantages of the full Darcy formulation are the increased number of degrees of freedom or the imposition of a condition on the normal velocity component of the boundary (impermeable domain). 
For a comparison of the full and reduced Darcy model in terms of solution time, memory requirements and accuracy we refer the interested reader to \cite{RefReducedDarcy}.

\section{Numerical examples}\label{sec:5}
In this section, we study numerical examples to demonstrate the behaviour of our modeling framework based on the connection of the poroelastic model and the synthetic vascular trees.

\subsection{Poroelastic circular disk coupled to planar trees}
We first 
consider a poroelastic circular domain that is perfused by a fluid provided by a planar supplying tree and returned into a planar draining tree (see also Fig. \ref{fig:coupledvesseltree}).
For the poroelastic disk, we choose the parameters in SI units summarized in Tab. \ref{table_simulation_parameters}. Instead of a spring-type condition, we fix the outer boundary for the moment, so that $\boldsymbol{u} = \boldsymbol{0}$ at the outer circular boundary. Each tree consists of 50 terminal vessels. For the bell-shaped source terms in (\ref{eqn:gauss}), we choose $b=3$.

\begin{table}
\caption{Model parameters for the poroelastic disk.}
\label{table_simulation_parameters}
\begin{tabular}{llll}
\hline
Skeleton-related parameters & & & Flow-related parameters  \\
\hline
Disk radius $r$ = 0.01 m & & & Initial porosity $\phi_0$ = 0.5  \\
Young's modulus $E$ = 1 $\frac{\text{kg}}{\text{m} \, \text{s}^2}$  & & &  Permeability $k$ =  $ 3.6 \cdot 10^{-3} \text{m}^2$ \\
Poisson's ratio $\nu$ = 0.3 & & & Dynamic viscosity $\eta = 3.6 \cdot 10^{-3}$ $\frac{\text{kg}}{\text{m} \,\text{s}}$ \\
& & & Perfusion flow (at root) $Q_\text{perf} = 800 \cdot 10^{-9} \frac{\text{m}^3}{\text{s}}$ \\[4pt] \hline
\end{tabular}
\vspace*{-4pt}
\end{table}

\begin{figure}[ht]
\centering
 \begin{tikzpicture}
   \node[] (pic) at (0,0) {\includegraphics[width=75mm]{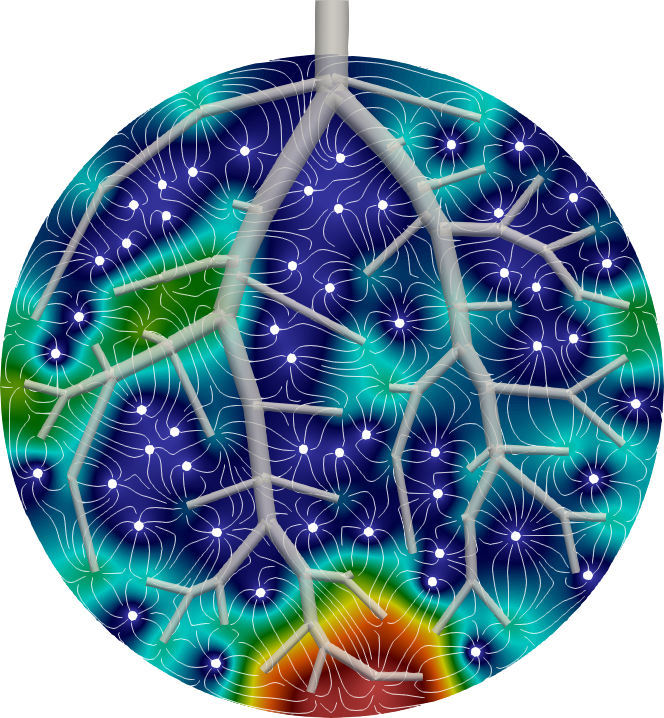}};
   \node[] (pic) at (5.0,-0.5) {\includegraphics[width=15mm]{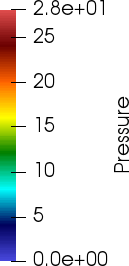}};
\end{tikzpicture}
\caption{Solution for the pressure field $p$ [$\frac{\text{kg}}{\text{m} \, \text{s}^2}$]. The depicted tree is the supplying tree. The streamlines illustrate flow from the inlets to the outlets.}
\label{fig:pressure_50}
\end{figure}

\begin{figure}[ht]
\centering
\subfigure[Displacement $u$\label{a}]{
 \begin{tikzpicture}
   \node[] (pic) at (0,0) {\includegraphics[width=50mm]{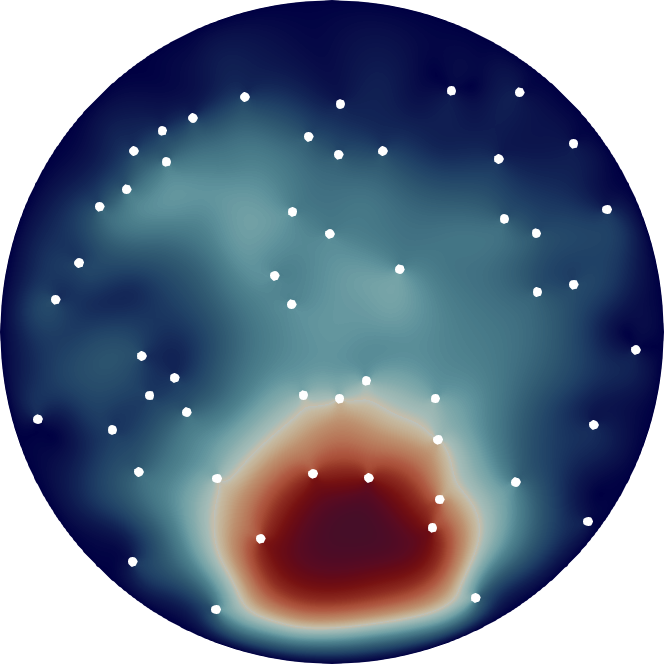}};
   \node[] (pic) at (3.25,0) {\includegraphics[width=10mm]{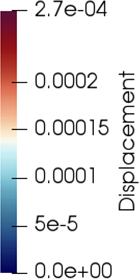}};
\end{tikzpicture}}
\hfill
\subfigure[Porosity $\phi$\label{b}]{
 \begin{tikzpicture}
   \node[] (pic) at (0,0) {\includegraphics[width=50mm]{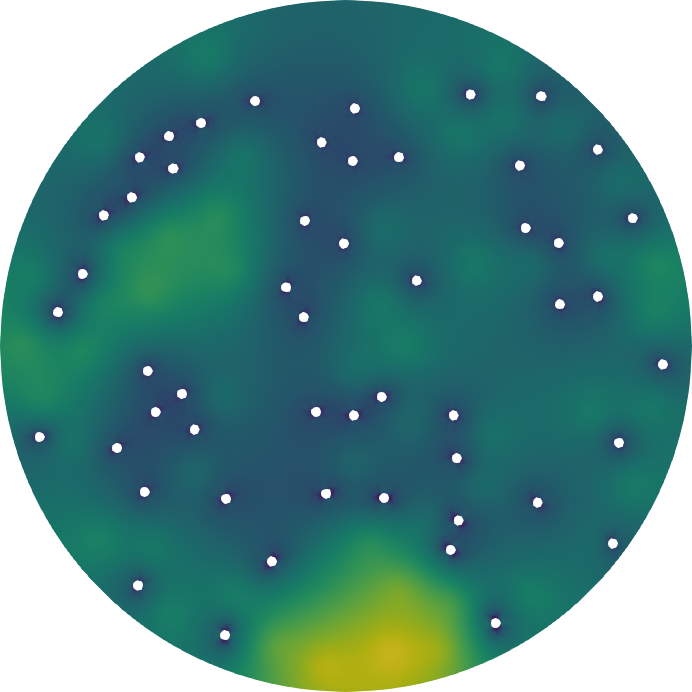}};
   \node[] (pic) at (3.25,0) {\includegraphics[width=10mm]{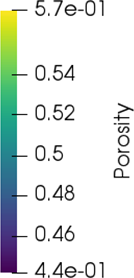}};
\end{tikzpicture}}
  \caption{Solution for displacement field $\boldsymbol{u}$ [m] with a fixed outer boundary and solution for porosity field $\phi$.}
\label{fig:displacement_porosity_50}
\end{figure}

We discretize the circular domain with a mesh of 36,826 triangular elements. We first obtain the solutions for the primary field variables $\boldsymbol{u}$ and $p$. With the pressure $p$ known, we can compute the porosity field $\phi$ from (\ref{eqn:pressureporosity}) and the velocity from Darcy's law (\ref{eqn:darcy}).
The solution of the pressure  $p$ is depicted in \ref{fig:pressure_50}. The white streamlines indicate the flow direction. 
One can observe higher pressure levels close to the inlet vessels. In areas without outlet points, the pressure reaches maximum values, such that a pressure gradient can be built up that drives the fluid to an outlet further away. 
The displacement solution $\boldsymbol{u}$ and the porosity field $\phi$ are plotted in Figs.  \ref{a} and \ref{b}. Higher displacement values can be observed in the areas of high pressure values. 
The porosity field $\phi$ fluctuates around the initial porosity value of 0.5. 



\subsubsection{Model sensitivity with respect to bell-shaped source term}
\begin{figure}[ht]
\centering
\subfigure[Solution for $b=1$ \label{a1}]{
 \begin{tikzpicture}
   \node[] (pic) at (0,0) {\includegraphics[width=50mm]{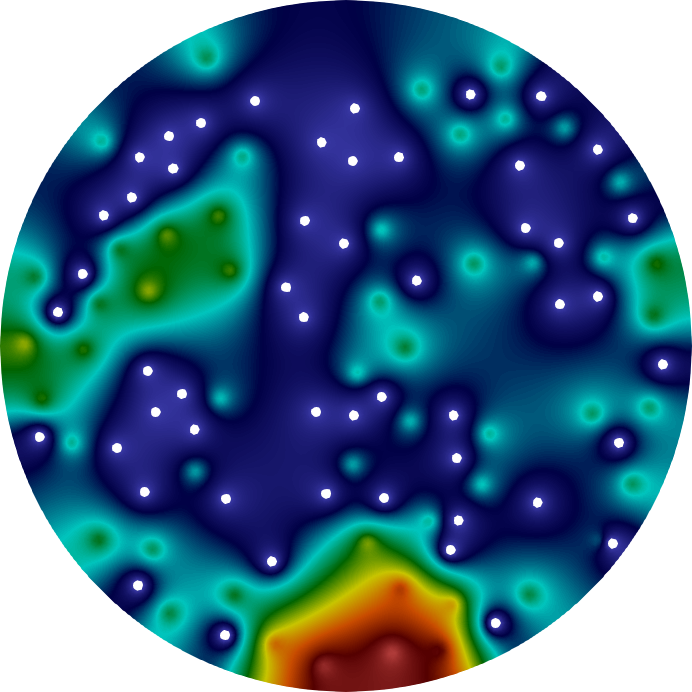}};
   \node[] (pic) at (3.25,0) {\includegraphics[width=10mm]{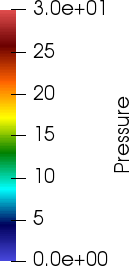}};
\end{tikzpicture}}
\hfill
\subfigure[Solution for $b = 3$ \label{b1}]{
 \begin{tikzpicture}
   \node[] (pic) at (0,0) {\includegraphics[width=50mm]{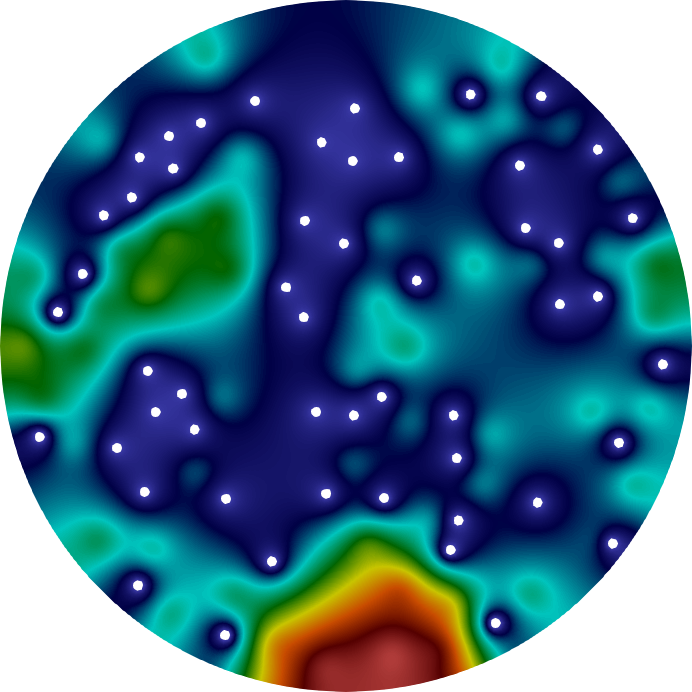}};
   \node[] (pic) at (3.25,0) {\includegraphics[width=10mm]{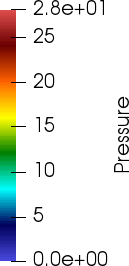}};
\end{tikzpicture}}
 \caption{Solution for pressure field $p$ [$\frac{\text{kg}}{\text{m} \, \text{s}^2}$] with different scaling factors $b$ of bell-shaped function.}
\label{fig:pressure_gauss}
\end{figure}
Figure \ref{fig:pressure_gauss} depicts the pressure field $p$ for two different values of the scaling factor $b$ of the bell-shaped function (\ref{eqn:gauss}). In the case of $b = 1$ (see Fig. \ref{a1}), the resulting pressure values are centered on a smaller area. Therefore, the maximum values also exceed the ones obtained with $b = 3$ (see Fig. \ref{b1}). Nevertheless, the global behaviour is in both cases equivalent. For all further computations, we proceed with $b = 3$.

\subsubsection{Model sensitivity with respect to stiffness of surrounding tissue}
\begin{figure}[ht]
\centering
\subfigure[Stiff contact with $\alpha = 5\cdot 10^2$ \label{a2}]{
 \begin{tikzpicture}
   \node[] (pic) at (0,0) {\includegraphics[width=50mm]{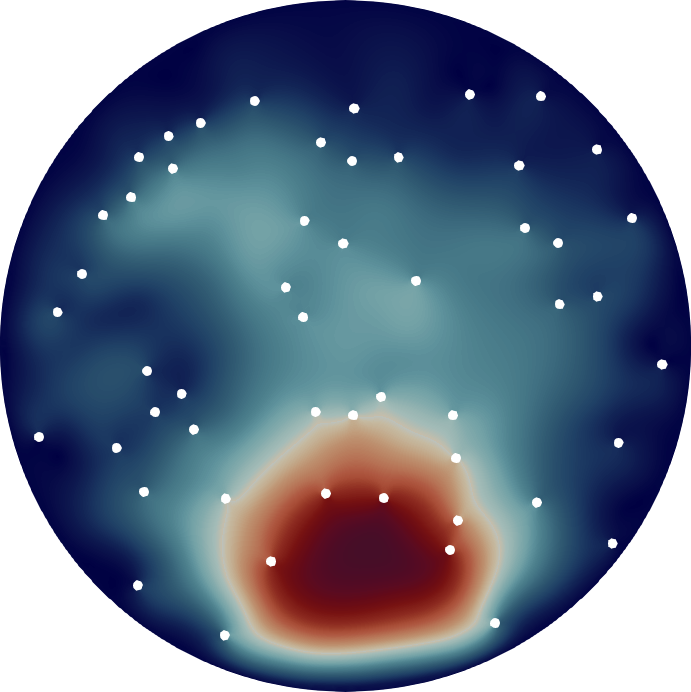}};
   \node[] (pic) at (3.25,0) {\includegraphics[width=10mm]{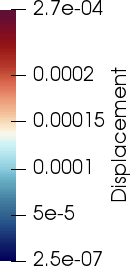}};
\end{tikzpicture}}
\hfill
\subfigure[Soft contact with $\alpha = 1\cdot 10^1$ \label{b2}]{
 \begin{tikzpicture}
   \node[] (pic) at (0,0) {\includegraphics[width=50mm]{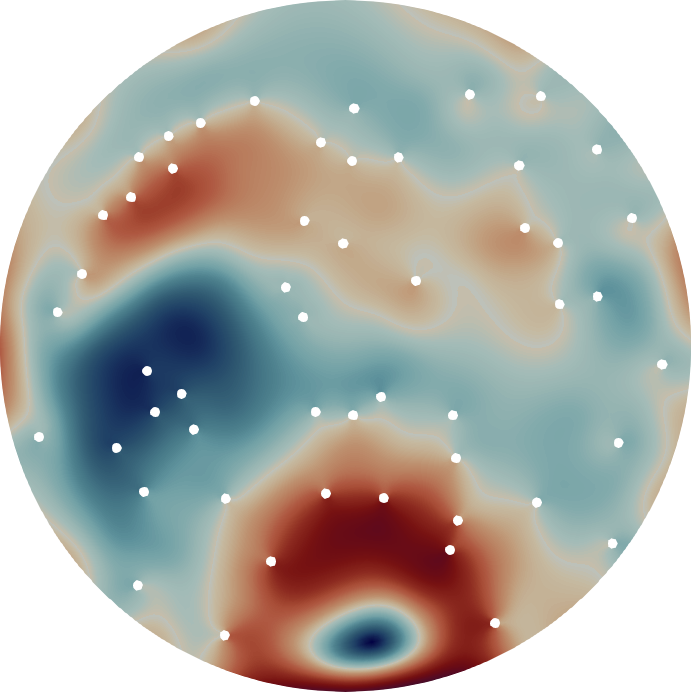}};
   \node[] (pic) at (3.25,0) {\includegraphics[width=10mm]{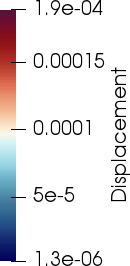}};
\end{tikzpicture}}
\caption{Solution for displacement field $\boldsymbol{u}$ [m] for different stiffness values in the contact boundary conditions.} 
\label{fig:stiffness_contact}
\end{figure}

Figure \ref{fig:stiffness_contact} plots the displacement solution for a stiffer ($\alpha = 5\cdot 10^2$) and softer resistance ($\alpha = 1\cdot 10^1$) in equation (\ref{eqn:stiffnesscontact}). In the stiff case (see Fig. \ref{a2}), the displacement field virtually indistinguishable from the solution with fixed boundary depicted in Fig. \ref{a}.
In the soft case, the boundary can deform, leading to a significantly different displacement pattern, plotted in Fig. \ref{b2}. Due to the weakening of the constraint in the soft case, the maximum displacement value decreases compared to the one in the stiff case.

\subsubsection{Model sensitivity with respect to hierarchical tree depth}
\begin{figure}[ht]
\centering
\subfigure[Pressure $p$ for 250 inlets/outlets \label{a3}]{
 \begin{tikzpicture}
   \node[] (pic) at (0,0) {\includegraphics[width=50mm]{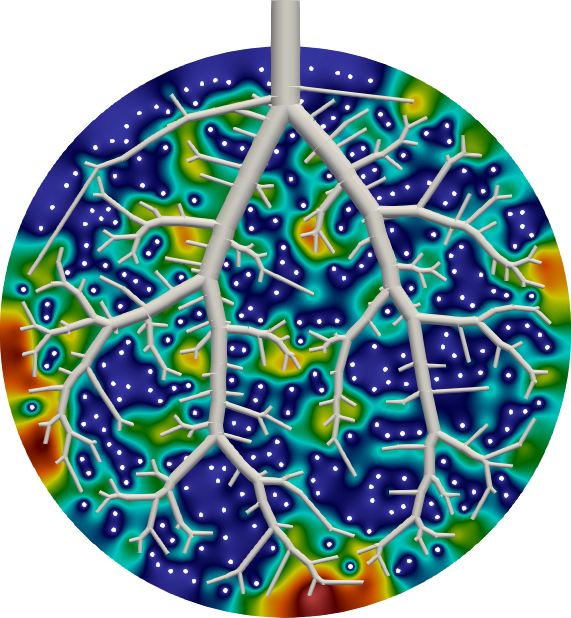}};
   \node[] (pic) at (3.25,-0.25) {\includegraphics[width=10mm]{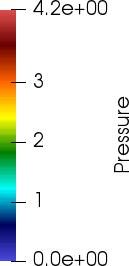}};
\end{tikzpicture}}
\hfill
\subfigure[Pressure $p$ for 1500 inlets/outlets \label{b3}]{
 \begin{tikzpicture}
   \node[] (pic) at (0,0) {\includegraphics[width=50mm]{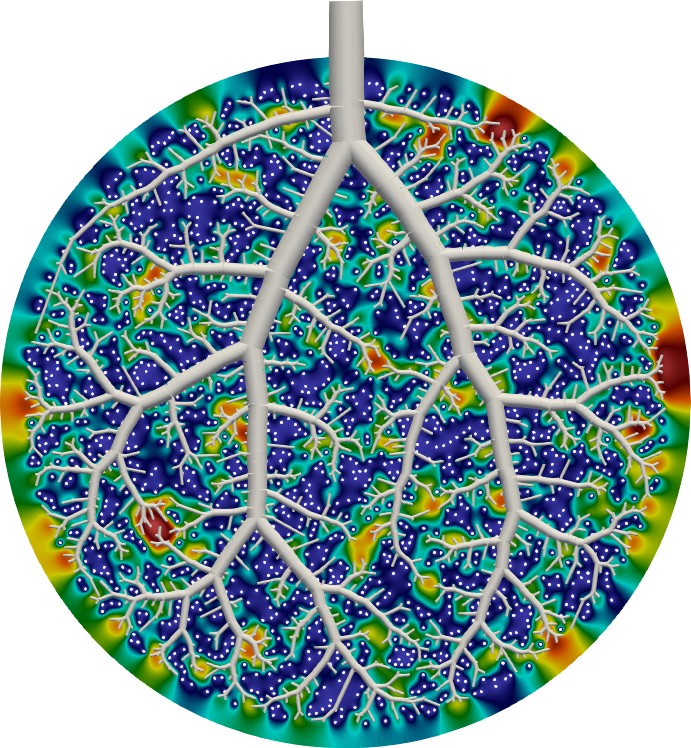}};
   \node[] (pic) at (3.25,-0.25) {\includegraphics[width=10mm]{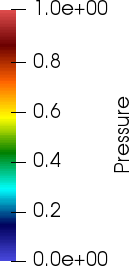}};
\end{tikzpicture}}
  \caption{Solution for pressure field $p$ [$\frac{\text{kg}}{\text{m} \, \text{s}^2}$] obtained with two different tree resolutions.}
  \label{fig:tree_hierarchies}
\end{figure}

We finally investigate the model behaviour at two different tree depths with 250 and 1,500 terminal vessels for both trees. In order to resolve the circular voids adequately, we refine the mesh with 114,320 elements in the former case to a mesh with 353,344 triangular elements in the latter case. 
The results of the pressure field $p$ are presented in Fig. \ref{fig:tree_hierarchies}. It is evident that the pressure drop between inlets and outlets is smaller in the case of a finer tree hierarchy (see Fig. \ref{a3} and \ref{b3}). When the trees are resolved with a larger depth, the outlet and inlet points seem to be more homogeneously distributed from a global perspective, resulting in shorter distances between the inlet and outlet. Thus, the global behaviour leads to a pressure solution that shows a more fine grained distribution. 
If we characterize a certain number of inlets or outlets with a representative volume element (RVE), we observe that the relative pattern of the solution with respect to such an RVE does not change. It is easy to verify from the plots that one can find similar patterns of the pressure field of the coarser tree in the pressure field of the finer tree. 

\subsection{Towards simulation based assessment of liver resection}
\label{sec:5b}

The liver has a unique ability to regenerate itself after damage. As a consequence, liver resections can be performed in which up to 75\% of the liver can be removed \cite{RefMichalopoulos}, for instance, to remove a cancerous tumor. A liver resection requires careful patient-specific planning in order to minimize the risk of liver failure. Since the liver is characterized by a high degree of vascularization, the regeneration process of the liver is dependent on the perfusion and redistributed flow after resection, which affects important functions such as blood supply or metabolism \cite{RefMichalopoulos}. During liver resection, the surgeon needs to consider various factors, such as the location and size of the tumor, the extent of liver tissue to be removed, and the preservation of the remaining liver tissue to maintain liver function.

\begin{figure}[ht]
\centering
 \begin{tikzpicture}
   \node[] (pic) at (0,0) {\includegraphics[width=50mm]{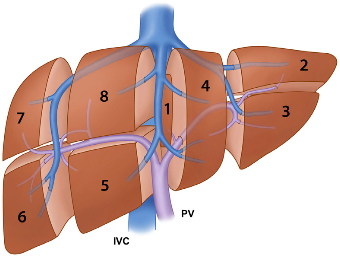}};
\end{tikzpicture}
\caption{Division of the human liver into eight segments corresponding to the portal vein and inferior vena cava (anterior view). From \cite{RefLiverSegments} (Licence: CC BY).}
  \label{fig:LiverSegments}
\end{figure}

In practice, there exists two ways of carrying out a resection. One is the anatomical resection, where one or more of the eight liver segments are removed. In that approach, the liver is divided into eight functionally independent segments which allow a resection of segments without damaging other segments \cite{RefVibert}. Each segment has its own supply by a larger vessel of the supplying tree that splits into smaller ones within the segment, and belongs to a branch of the draining tree (see Fig. \ref{fig:LiverSegments}). The second option is a non-anatomical cut which takes place when a tumor is distributed over many segments and a bigger portion of tissue needs to be removed. In this case, the surgeon is faced with the decision between the risk of tumor recurrence and the risk of liver failure \cite{RefChrist}. 

In this context, understanding the redistributed flow and mechanical response, e.g. stress or pressure accumulation areas, after resection has clinical relevance \cite{RefMichalopoulos}. In the worst case, a cut might cause so-called orphans which are parts of the vessels trees that are not supplied with blood anymore. 
In the following, we will employ our modeling framework to evaluate cut patterns and investigate blood flow redistribution after surgical resection.


\subsubsection{Patient-specific liver geometry and discretely resolved vascularization}
\begin{figure}[ht]
\centering
\subfigure[2D slice \label{a4}]{
 \begin{tikzpicture}
   \node[] (pic) at (0,0) {\includegraphics[width=62mm]{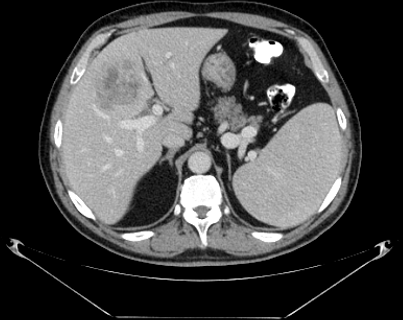}};
\end{tikzpicture}}
\hfill
\subfigure[Segmentation mask with liver in green \label{b4}]{
 \begin{tikzpicture}
   \node[] (pic) at (0,0) {\includegraphics[width=62mm]{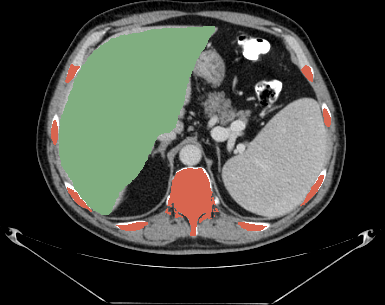}};
\end{tikzpicture}}
    \caption{Abdominal CT scan containing the liver with a resolution of 0.977 x 0.977 mm and a slice thickness of 2.5 mm.} 
  \label{fig:segmentation}
\end{figure}
We generate a patient-specific liver model based on imaging data obtained from CT scans \cite{RefSegmentation}. 
For the segmentation of the liver, we use the open source software package 3D Slicer\footnote[1]{https://www.slicer.org/} and the free software Autodesk Meshmixer\footnote[2]{https://meshmixer.com/}.
A 2D slice of the 3D voxel model and the segmentation mask of the liver domain (green colour) are shown in Figs. \ref{a4}. and \ref{b4}, respectively. The resolution of the CT scan is 0.977 x 0.977 mm within each image, with a spacing of 2.5 mm between the slices.

\begin{figure}[ht]
\centering
\subfigure[Anterior view \label{a5}]{
 \begin{tikzpicture}
   \node[] (pic) at (0,0) {\includegraphics[width=55mm]{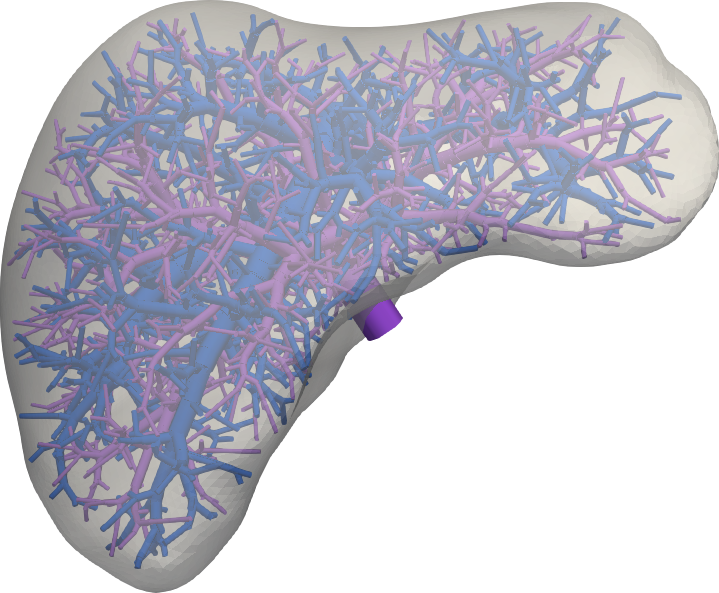}};
\end{tikzpicture}}
\hfill
\subfigure[Inferior view \label{b5}]{
 \begin{tikzpicture}
   \node[] (pic) at (0,0) {\includegraphics[width=55mm]{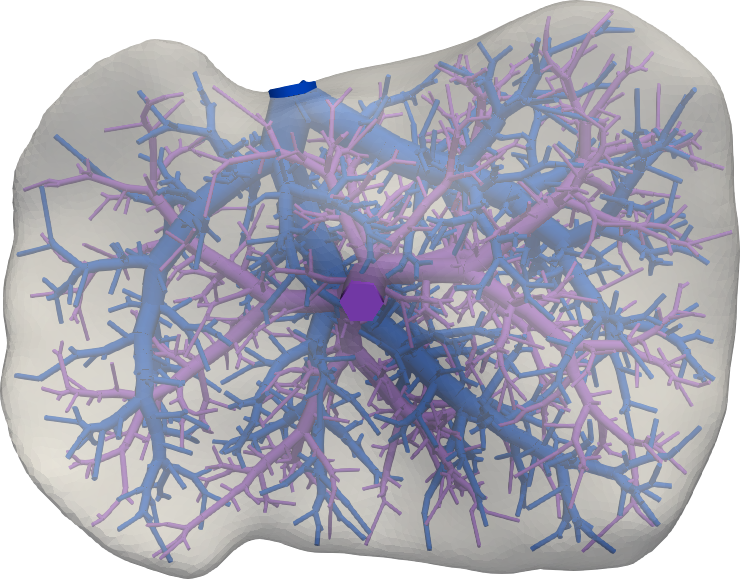}};
\end{tikzpicture}}
  \caption{Patient-specific 3D liver model with the supplying (purple) and draining (blue) vascular tree structures.} 
  \label{fig:LiverModelWithVessels}
\end{figure}

As the hepatic artery and portal vein are mostly aligned, they are usually combined in one single tree for simplicity 
\cite{Jessen3}.
Figure \ref{fig:LiverModelWithVessels} illustrates the segmented liver with the synthetic supplying tree (hepatic artery and portal vein) and the synthetic draining tree (hepatic vein). For both trees we choose 1,000 terminal vessels to model the flow of blood into the poroelastic domain in an accurate manner, while still maintaining computational efficiency. We note that we recently improved the efficiency of the vascular generation algorithm described in section \ref{sec:3} , which allows us to generate full scale vascular trees with around 1,000,000 terminal vessels \cite{RefEtienneMurray}.

After creating the liver geometry, we assume spherical voids at the terminal points of the outlets where we impose zero pressure as a reference level. We then generate a mesh which contains 7,385,996 tetrahedral elements.

\subsubsection{Anatomical vs. non-anatomical resection}
\begin{figure}[ht]
\centering
\subfigure[Velocity \label{a8}]{
 \begin{tikzpicture}
   \node[] (pic) at (0,0) {\includegraphics[width=50mm]{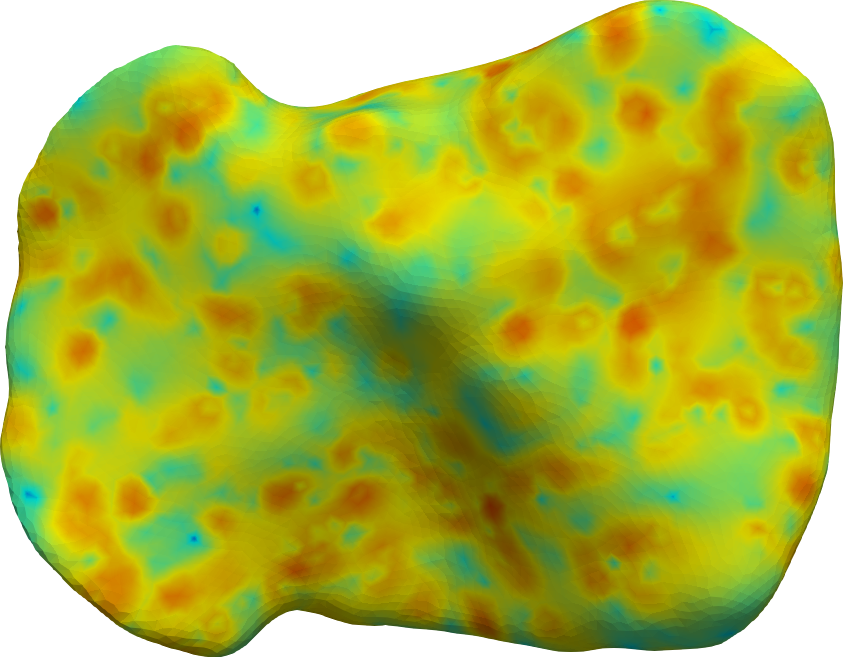}};
      \node[] (pic) at (3.25,0) 
      {\includegraphics[width=10mm]{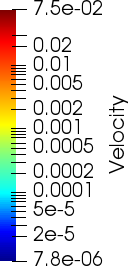}};
\end{tikzpicture}}
\hfill
\subfigure[Pressure \label{b8}]{
 \begin{tikzpicture}
   \node[] (pic) at (0,0) {\includegraphics[width=50mm]{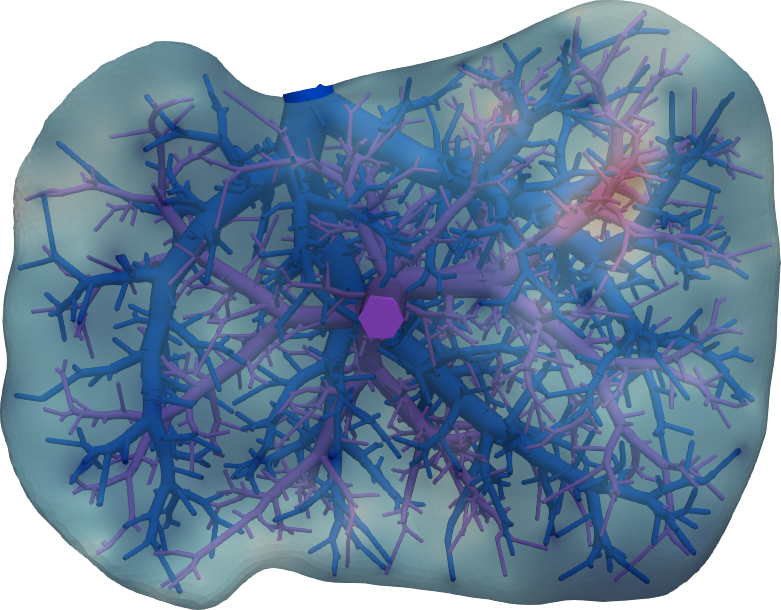}};
   \node[] (pic) at (3.25,0) 
   {\includegraphics[width=10mm]{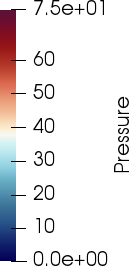}};
\end{tikzpicture}}
  \caption{Solution for velocity $\boldsymbol{w}$ [$\frac{\text{m}}{\text{s}}$] and pressure field $p$ [$\frac{\text{kg}}{\text{m} \, \text{s}^2}$] of full liver model (inferior view).}
  \label{fig:results_full_liver}
\end{figure}

\begin{table}
\caption{Simulation parameters for the liver problem \cite{RefLiverPerfusion,RefDebbautMaterialproperties,RefElasticPropertiesLiver}.}
\label{table_simulation_parameters_liver}
\begin{tabular}{llll}
\hline
Tissue deformation-related parameters & & & Perfusion-related parameters\\
\hline  \\[-10pt]
Young's modulus $E$ = 5000 $\frac{\text{kg}}{\text{m} \, \text{s}^2}$ & & & Initial porosity $\phi_0$ = 0.15 \\
 Poisson's ratio $\nu$ = 0.35 & & &  Permeability $k$ = $2\cdot 10^{-14}$ $\text{m}^2$ \\
 & & & Dynamic viscosity $\eta = 3.6 \cdot 10^{-3}$ $\frac{\text{kg}}{\text{m} \,\text{s}}$
 \\
& & & Inflow (at root) $Q_\text{perf} = 20\cdot10^{-6} \frac{\text{m}^3}{\text{s}}$ \\[4pt] \hline
\end{tabular}
\vspace*{-4pt}
\end{table}
Detecting areas with insufficient blood supply and locally quantifying the perfusion efficiency is helpful for the assessment of the post-operative outcome. 
We first show the results of the liver model before resection. The physiological parameters that have been used for all liver computations are listed in Tab. \ref{table_simulation_parameters_liver}.
The results for pressure and velocity are depicted in Fig. \ref{fig:results_full_liver}. As can be seen in \ref{a8}, the unresected model shows a homogeneous blood supply to the liver tissue, which can be expected in a healthy liver state. Also the pressure field in Fig. \ref{b8} does not show disparities or areas of pressure accumulation.

\begin{figure}[ht]
\centering
   \begin{tikzpicture}
   \node[] (pic) at (0,0) {\includegraphics[width=75mm]{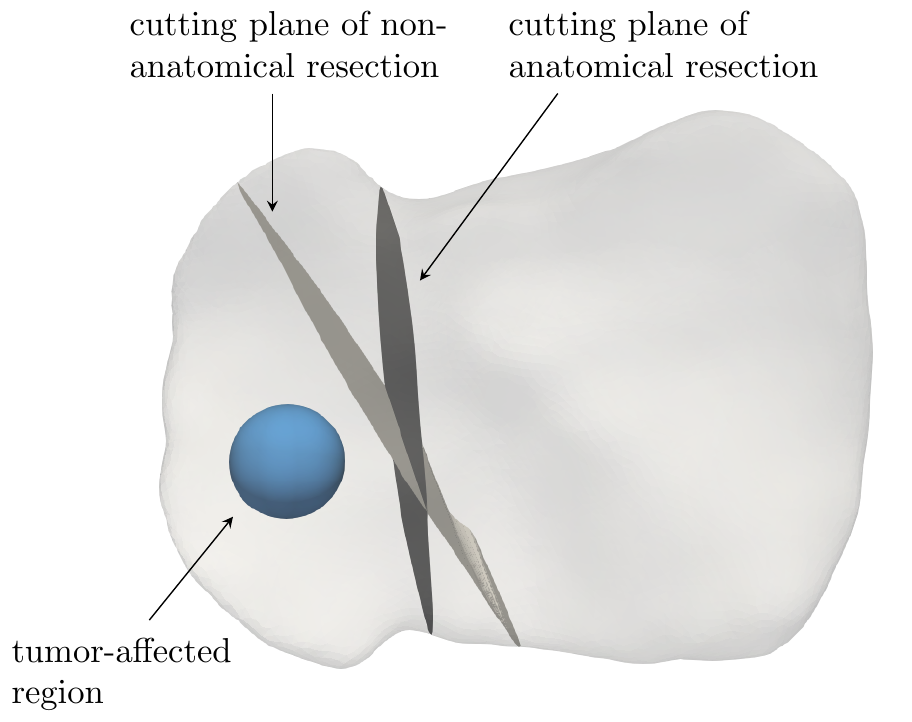}};
     \end{tikzpicture}
 	\caption{Two cutting planes for the resection of liver tissue. The blue sphere represents the tumor-affected region.}
	\label{fig:cuttingplanes}
\end{figure}

\begin{figure}[ht]
\centering
\subfigure[Anatomical resection of the left lateral section (segments 2 and 3) \label{a6}]{
 \begin{tikzpicture}
   \node[] (pic) at (0.0,0) {\includegraphics[width=40mm]{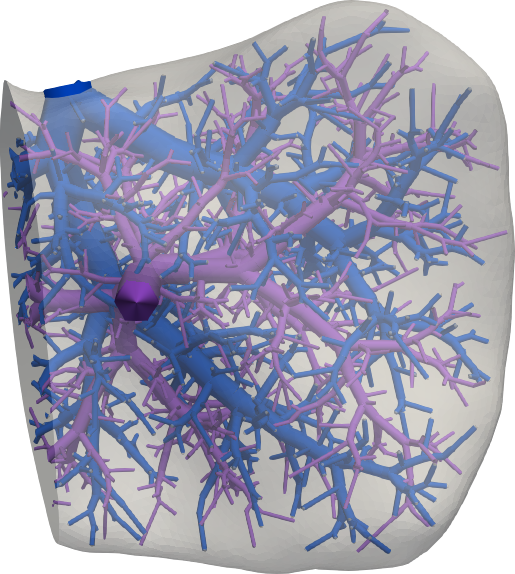}};
       \node[opacity=0.0] (pic) at (2.75,0) {\includegraphics[width=10mm]{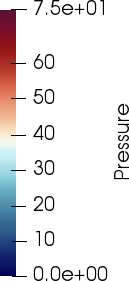}};
\end{tikzpicture}}
\hfill
\subfigure[Non-anatomical resection \label{b6}]{
 \begin{tikzpicture}
   \node[] (pic) at (0.0,0) {\includegraphics[width=51mm]{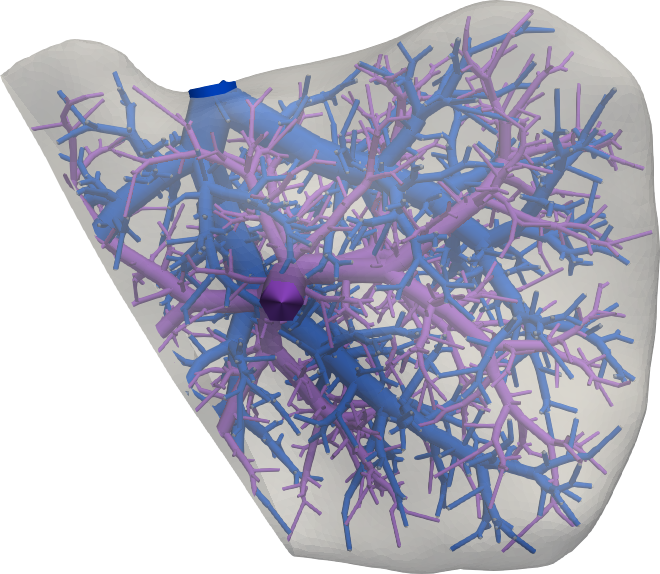}};
       \node[opacity=0.0] (pic) at (2.3,0) {\includegraphics[width=10mm]{"./Images/pressure_full_colorbar".png}};
\end{tikzpicture}}
  \caption{Model representation of the resected liver.}
  \label{fig:resection_models}
\end{figure}

We now assume that the left lateral section of the liver is affected by a tumor. We use our framework for modeling perfusion to investigate the behaviour of the liver after resection. In particular, we consider two options for potential cuts that are illustrated in Fig. \ref{fig:cuttingplanes}. Figure \ref{fig:resection_models} illustrates the remaining domain of the liver and the remaining vascular tree after resection for both cut options in the inferior view. The first cut option in \ref{a6} corresponds to an anatomical resection of the left lateral section in which the liver segments 2 and 3 are removed (see Fig. \ref{fig:LiverSegments}). The discretization of the remaining liver domain after anatomical resection consists of a mesh with 5,882,171 tetrahedral elements. The second cut option in \ref{b6} corresponds to a non-anatomical resection with a diagonal cut. The discretization of the remaining liver domain consists of a mesh with 6,107,676 tetrahedral elements.

We note that vessels resolved in the vascular tree structure which are cut must be closed during surgery to prevent blood loss. In our simulations, we therefore do not allow blood flow through any vessel that is cut, and the blood flow of all cut vessels is redistributed over the remaining portion of the intact tree.



\begin{figure}[ht]
\centering
\subfigure[Anatomical resection \label{a7}]{
 \begin{tikzpicture}
   \node[] (pic) at (0.0,0) {\includegraphics[width=40mm]{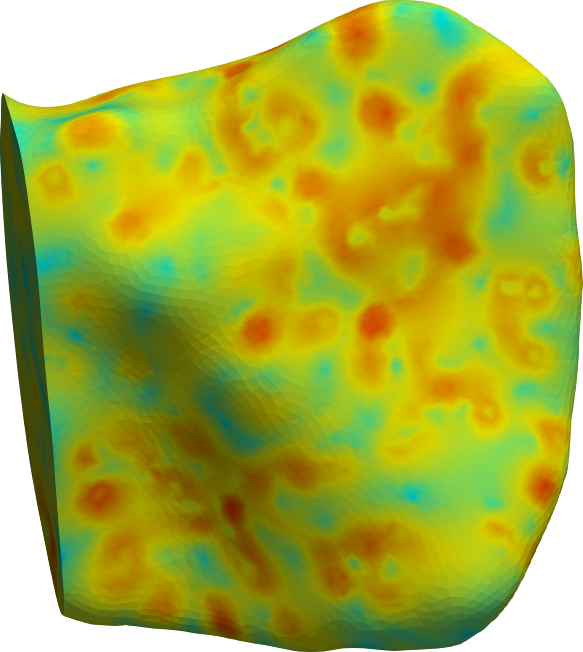}};
    \node[] (pic) at (2.75,0) 
     {\includegraphics[width=10mm]{"./Images/Liver_velocity_new_colorbar".png}};
\end{tikzpicture}}
\hfill
\subfigure[Non-anatomical resection \label{b7}]{
 \begin{tikzpicture}
    \node[] (pic) at (0.0,0) {\includegraphics[width=60mm]
    {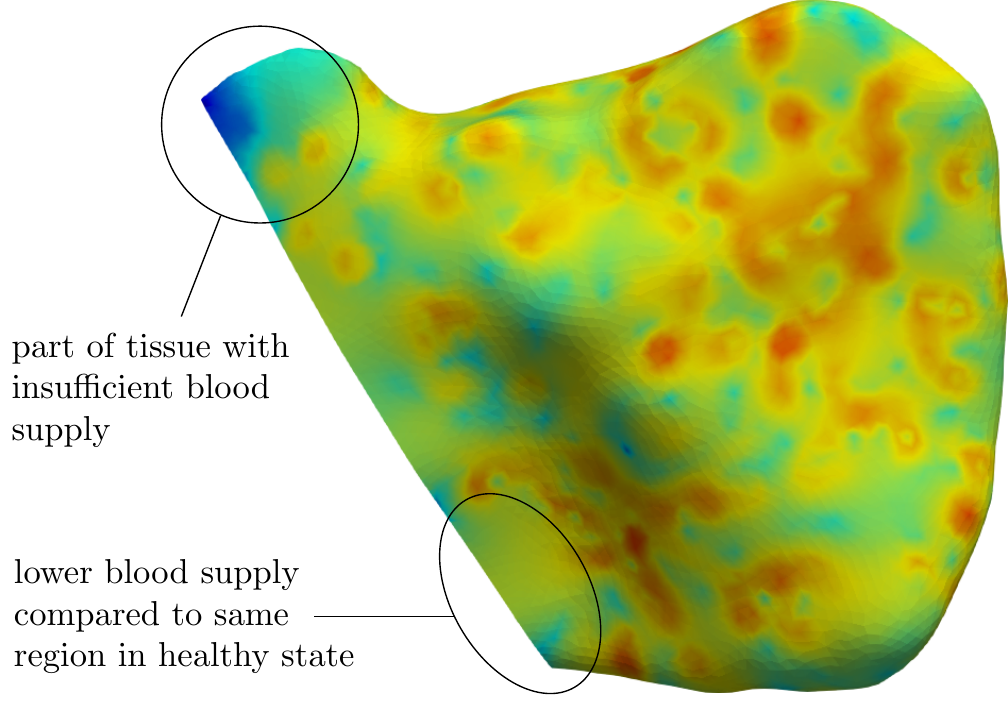}};
    \node[] (pic) at (3.75,0)
     {\includegraphics[width=10mm]{"./Images/Liver_velocity_new_colorbar".png}};
\end{tikzpicture}}
  \caption{Solution for velocity $\boldsymbol{w}$ [$\frac{\text{m}}{\text{s}}$] after resection.}
  \label{fig:velocity_resection_models}
\end{figure}

The simulation results, shown in Fig. \ref{fig:velocity_resection_models}, clearly outline the difference in blood supply for the two cuts. While the anatomical resection in Fig. \ref{a7} causes a homogeneously distributed perfusion of the domain, the non-anatomical resection in Fig. \ref{b7} leads to a part of liver tissue with insufficient blood supply and a part of tissue with lower blood supply compared to the same region in the unresected liver shown in Fig. \ref{a8}. We hence conclude that the diagonal cut would suffer from uneven blood supply in the post-operative regenerative process.

\begin{figure}[ht]
\centering
\subfigure[Anatomical resection \label{a9}]{
 \begin{tikzpicture}
   \node[] (pic) at (0.0,0) {\includegraphics[width=40mm]{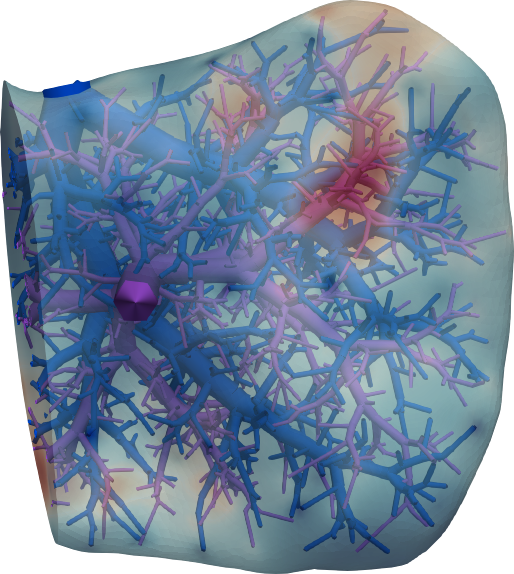}};
    \node[] (pic) at (2.75,0) 
     {\includegraphics[width=10mm]{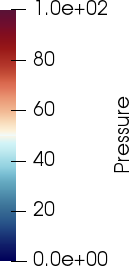}};
\end{tikzpicture}}
\hfill
\subfigure[Non-anatomical resection \label{b9}]{
 \begin{tikzpicture}
   \node[] (pic) at (0.0,0) {\includegraphics[width=51mm]{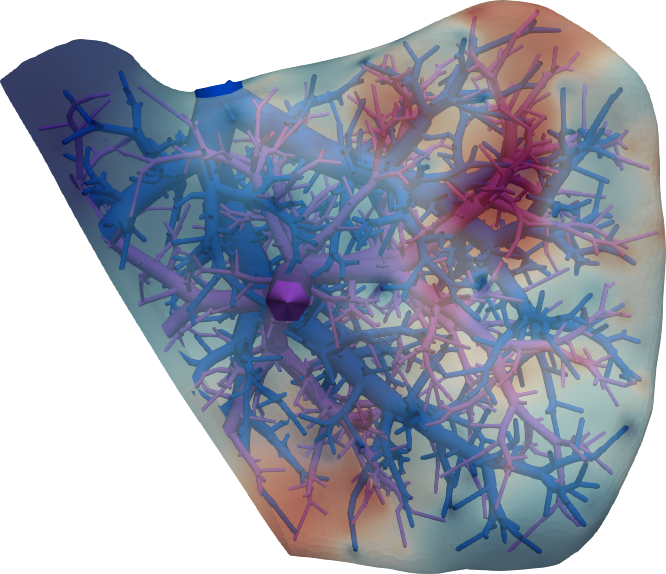}};
    \node[] (pic) at (3.3,0) 
     {\includegraphics[width=10mm]{"./Images/Liver_pressure_resection_new_colorbar".png}};
\end{tikzpicture}}
 \caption{Solution for pressure field $p$ [$\frac{\text{kg}}{\text{m} \, \text{s}^2}$] after resection.}
  \label{fig:pressure_resection_models}
\end{figure}

In Fig. \ref{fig:pressure_resection_models}, we compare the corresponding pressure fields. We observe that both cut options lead to higher pressure levels in the liver after resection compared to the unresected liver shown in Fig. \ref{b8}. This phenomenon is physiological and known as hyperperfusion. It occurs because the same amount of blood must now pass through a smaller remaining liver domain.  Moreover, the non-anatomical resection in Fig. \ref{b9} exhibits more areas with pressure accumulation (plotted in red) and higher disparities in the pressure distribution than the anatomical resection in \ref{a9}.

\section{Discussion and outlook}\label{sec:6}
In this paper, we presented a modeling framework that connects continuum poroelasticity and discrete vascular tree structures to model liver tissue in terms of perfusion and deformation. The connection is achieved through a series of modeling assumptions and decisions. Firstly, we used bell-shaped functions as source terms in the pressure equation to impose inflow at the interfaces of the terminal vessels of the supplying tree and the poroelastic domain. Secondly, we introduced void regions that model the interface between the terminal vessels of the draining tree and the poroelastic domain, where pressure boundary conditions could be applied accordingly. Additionally, we took into account contact to surrounding tissue, using nonlinear springs at the boundary of the poroelastic domain. We demonstrated the numerical behaviour and versatility of our modeling framework via a poroelastic circular disc connected to planar trees. We performed a series of sensitivity studies to test the model behaviour with respect to source term parameters, stiff and soft contact and hierarchical tree depth. 

We then investigated our modeling framework for a realistic liver problem that consisted of two different resection scenarios of a patient-specific liver, one anatomical with an expected satisfactory results and one non-anatomical with an expected non-satisfactory result. We showed how patient-specific data can be incorporated into our model and then computed the flow redistribution after the two different cuts. As expected, the numerical results indicate a difference in blood supply for the two resection scenarios, in which the anatomically resected liver performed satisfactorily and the non-anatomically resected liver exhibited parts with insufficient blood supply.

In summary, our results demonstrate that the combination of poromechanics and synthetic vascular trees can enable useful and accurate tools for modeling liver tissue. Although a robust validation study is still lacking, we can already observe that the presented approach has the potential to aid in assessing and optimizing surgical treatment procedures. 
In this sense, the model presented here constitutes another step towards patient-specific evidence-based physiological  simulation tools that can be applied in clinical practice. In order to fully ensure applicability, material parameters such as elastic parameters, permeability and porosity must be further personalized, and a number of validation studies need to be performed. These aspects are subject of ongoing work.

In addition, we think that the model must be further refined, potentially driven trough future results from validation studies. One idea is to complement the poroleastic model by multiple compartments. Instead of lumping the lower levels of the vascular tree and the microcirculation together, compartmentalized poroelasticity would allow us to represent perfusion and deformation within the lower levels of the perfusion tree and the microcirculation separately, taking into account their different physiology. 
Furthermore, deformation and stresses play an important role for the further development towards modeling liver regrowth after surgical resection. Currently, we extend our framework by a liver regrowth model on a patient-specific basis that is guided by the goal of reducing stresses in the liver tissue. In this context, we also go from the current one-way coupling without any reverse influence on the synthetic trees to a fully coupled model that takes into account deformation induced change of the location of the terminal vessels and their interface regions with the poroelastic domain.

\section*{Acknowledgment}
The results presented in this work were achieved as part of the ERC Starting grant project $\mathrm{'}$ImageToSim$\mathrm{'}$ that has received funding from the European Research Council (ERC) under the European Union’s Horizon 2020 research and innovation programme (grant agreement no. 759001). The authors gratefully acknowledge this support. The authors also gratefully acknowledge the computing time provided to them on the high-performance computer Lichtenberg at the NHR Centers NHR4CES at TU Darmstadt. This is funded by the Federal Ministry of Education and Research and the State of Hesse.

\end{document}